\documentclass[twocolumn]{aastex631}
\usepackage{tabularx}
\usepackage{enumitem}
\usepackage{makecell} 
\usepackage{hyperref}

\usepackage{tabularx}
\usepackage{array}
\newcolumntype{Y}{>{\centering\arraybackslash}X}

\usepackage{amsmath,amssymb}

\newcommand{\bham}{Institute for Gravitational Wave Astronomy and School of Physics and Astronomy, University of Birmingham, Edgbaston, Birmingham
B15 2TT, UK}
\newcommand{\qf}{Quantum Formatics, Cambridge, MA 02139, USA}
\newcommand{\wvu}{Department of Physics and Astronomy, West Virginia University, P.O. Box 6315, Morgantown, WV 26506, USA}
\newcommand{\gwac}{Center for Gravitational Waves and Cosmology, West Virginia University, Chestnut Ridge Research Building, Morgantown, WV 26505, USA}
\newcommand{\IISER}{Department of Physics, Indian Institute of Science Education and Research Bhopal, Bhopal Bypass Road, Bhauri, Bhopal 462066, Madhya Pradesh, India}

\begin{document}

\title{The Evolution of Nulling in Pulsars}

\author[0000-0003-3053-6538]{P.\,R.\,Brook}\affiliation{\bham}
\author[0000-0001-7974-5264]{J.\,B.\,Gibson}\affiliation{\qf}
\author[0000-0001-7697-7422]{M.\,A.\,McLaughlin}\affiliation{\wvu}\affiliation{\gwac}
\author[0000-0002-9507-6985]{M.\,P.\,Surnis}\affiliation{\IISER}
    
\begin{abstract}
Nulling is a phenomenon where the emission from a pulsar becomes undetectable (or significantly weaker) for a relatively short period of time, followed by a return to a normal emission state. The timescale of nulling ranges from a few pulse periods to many hours or even days. The fraction of
time a nulling pulsar spends in a null state varies across the population of canonical pulsars, from 0 to 95 per cent. The
long-term behaviour of a pulsar's nulling fraction, however, is currently unknown, as published values have typically been obtained through single observations. Here, we present the first long-term analysis of nulling behaviour in eight pulsars observed in the Parkes Multibeam Pulsar Survey over the course of eight to ten years. We also apply a new Bayesian method for pulse-energy analysis, yielding posterior estimates of the nulling fraction per observation. In several cases, the nulling affects only specific components of the pulse profile, rather than the entirety of the emission. Our analysis reveals that, while most pulsars show no significant trend in their nulling fraction over time, a subset exhibit some evidence for non-zero gradients in nulling fraction. In particular, PSRs J1048$-$3832, J1745$-$3040, and J1825$-$0935 show statistically significant trends over the span of the data. Studying the behaviour of nulling over years and decades is valuable as it can provide insights into the physical emission processes within pulsars. Studying how nulling evolves also provides valuable insights into pulsar evolution and the characterisation of the broader pulsar population.
\end{abstract}
 

\section{Introduction}
\label{sec:introduction}
\par{A pulsar is a highly magnetized, rotating neutron star that emits beams of electromagnetic radiation from its magnetic poles. When the Earth is enveloped by one of these sweeping beams, a short burst of radiation reaches our telescopes. As a result, we observe pulses at intervals set by the pulsar’s rotation rate. For most pulsars this is one pulse per rotation, and this interval (typically milliseconds to seconds) defines the \emph{pulse period}. Taking the average of many (typically thousands)  of such electromagnetic pulses results in a \emph{pulse profile}, a generally steady fingerprint of a pulsar's emission characteristics.
A few years after the discovery of pulsars, their largely predictable emission behaviour was occasionally observed to significantly deviate on short timescales, in the forms of nulling and mode changing \citep{1970Natur.228...42B, 1970Natur.228.1297B}. Mode changing is a phenomenon in which pulsars are seen to discretely switch between two or more emission states. Nulling can be thought of as an extreme form of mode changing, with one state showing little to no emission. The timescale of mode changing and nulling ranges from a few pulse periods to many hours or even days \citep{2007MNRAS.377.1383W}.
A related class of sources, the \emph{intermittent pulsars} \citep[e.g.,][]{2006Sci...312..549K}, show quasi-periodic switches between detectable and undetectable emission on timescales of days to years. This behaviour is sometimes described as long-timescale nulling.
The fraction of time in which a nulling pulsar is in a null state (the nulling fraction; NF) varies from 0 to 95 percent, and has been found to positively correlate with both characteristic age \citep{1976MNRAS.176..249R,2007MNRAS.377.1383W} and pulse period \citep{1976MNRAS.176..249R,1992ApJ...394..574B,10.1093/mnras/stab282}. However, other studies have found no significant correlation between NF and either characteristic age \citep{1986ApJ...301..901R,10.1093/mnras/stab282} or pulse period \citep{2007MNRAS.377.1383W}, suggesting that the relationship between nulling behaviour and pulsar parameters may be complex.

Our understanding of the radio emission mechanism is currently incomplete, but a pulsar's radiation beams are thought to result from intense magnetic fields at its magnetic poles and the charged particle currents that these fields produce. The phenomena of moding changing and nulling are believed to result from changes and failures in these particle currents.

At present, we are agnostic as to whether a particular pulsar's nulling or mode changing behaviour evolves over years and decades; published moding-changing and NFs have typically been obtained from single, long-duration observations, e.g., approximately two hours in \citet{2007MNRAS.377.1383W}.
Although the aforementioned intermittent pulsars are monitored on long timescales, accurately measuring their NFs, or tracking their evolution, would require high-cadence observations across their full on–off cycles.

Observing how pulsar phenomena evolve over many years has important implications for the understanding of physical processes intrinsic to the pulsar and its immediate environment. Such studies can also contribute to our understanding of pulsar evolution and the characterisation of the pulsar population as a whole. By investigating nulling behaviour, we can gain insight into the evolution of pulsars as revealed through changes in their emission properties. For example, if the frequency or duration of nulling events increases with time, it raises the possibility that these pulsars may transition into other types of intermittent emitters, such as Rotating Radio Transients \citep[RRATs;][]{2006Natur.439..817M}. Understanding these transitions is crucial for determining how many pulsars may go undetected due to their sporadic emission. Studying nulling pulsars, therefore, allows us to improve estimates of the Galactic pulsar population, refine population models, and gain insight into the evolutionary pathways that lead to different observational characteristics.

We analyse data from the Parkes Multibeam Pulsar Survey \citep[PMPS;][]{2001MNRAS.328...17M}, which spans 16 years and so allows us to begin to probe decadal timescales. To this end, our study focuses on eight pulsars that are either nulling pulsars or multi-component pulsars with a pulse-profile component that nulls. For each, we calculate NF values over approximately a decade.


The rest of this paper is organized as follows. In Section~\ref{sec:data} we detail the data used in our analysis. The methods we use to compute NFs are described in Section~\ref{sec:method}. In Section~\ref{sec:simulations} we test the efficacy of these methods using simulated nulling pulsar data. Section~\ref{sec:pulsars} describes the eight pulsars for which NF evolution is calculated in this work. The results of our analysis are presented in Section~\ref{sec:results}, discussed in Section~\ref{sec:discussion} and conclusions are drawn in Section~\ref{sec:conclusions}.}

\section{Data}
\label{sec:data}
We conduct our analyses on data from the PMPS, which began in 1997 and was conducted by the Parkes 64-m radio telescope. Employing a multibeam receiver comprising a hexagonal grid of 13 independent circular radio beams, the PMPS captured data across a 300~MHz radio bandwidth at a central frequency of 1374~MHz. Focused on a narrow section of the southern Galactic plane region ($|b| < 5^\circ$ and $260^\circ < l < 50^\circ$), the PMPS achieved unprecedented spatial coverage and sensitivity.

To identify pulsars exhibiting nulling behaviour within the PMPS, either in their overall emission or in specific components of their pulse profiles, we took two approaches: (i) leveraging nulling pulsars catalogued by the High Time Resolution Universe (HTRU) Pulsar Survey \citep[][]{burke2012high} and (ii) conducting direct scrutiny of the PMPS dataset. Eight pulsars were ultimately selected for this nulling analysis based on several additional criteria beyond evidence of nulling behaviour; selection was also influenced by the abundance and quality of a pulsar's observational data, and by how frequently it had been observed over long timescales. Pulsars prioritised for long-term monitoring, such as young or high-magnetic-field pulsars, were more likely to have been observed many times, providing a rich dataset for long-term nulling analysis.

The number of phase bins is set to 512 to provide a sufficient signal-to-noise ratio (S/N) for conducting the nulling analysis while maintaining enough resolution to preserve key pulsar features. Each pulsar is dedispersed and the time series is folded with the known period of the pulsar from the timing ephemeris. Any poor quality individual observations were manually excluded from the final dataset. This includes a small number of epochs exhibiting anomalous baseline distortions, which are discussed in detail for PSR~J1745$-$3040 in Section~\ref{sec:discussion} and Appendix~\ref{app:removed}.


\section{Calculating Nulling Fraction}
\label{sec:method}
\par{Unless the S/N is high, calculating the NF of a pulsar observation is not trivial. In reality, observations of nulling pulsars often contain single pulses that could be classified either as a non-nulling pulse with low S/N or as a noisy null pulse. Many formal methods for the calculation of an observation's NF are variations of a technique developed by \citet{1976MNRAS.176..249R}. We will refer to this as the \emph{histogram-scaling} (HS) technique and discuss it in detail in Section~\ref{sec:histogram_scaling}. We have also developed an alternative NF-calculation technique using Bayesian parameter estimation (BPE). This will be discussed in detail in Section~\ref{sec:bayesian_parameter_estimation}. Before an NF can be calculated (by either method), the data must undergo some additional processing, described in the following.}
\subsection{Data Preparation}
\label{sec:analysis_prep}
\par
\begin{figure*}
\centering
\includegraphics[width = 1.0\textwidth]{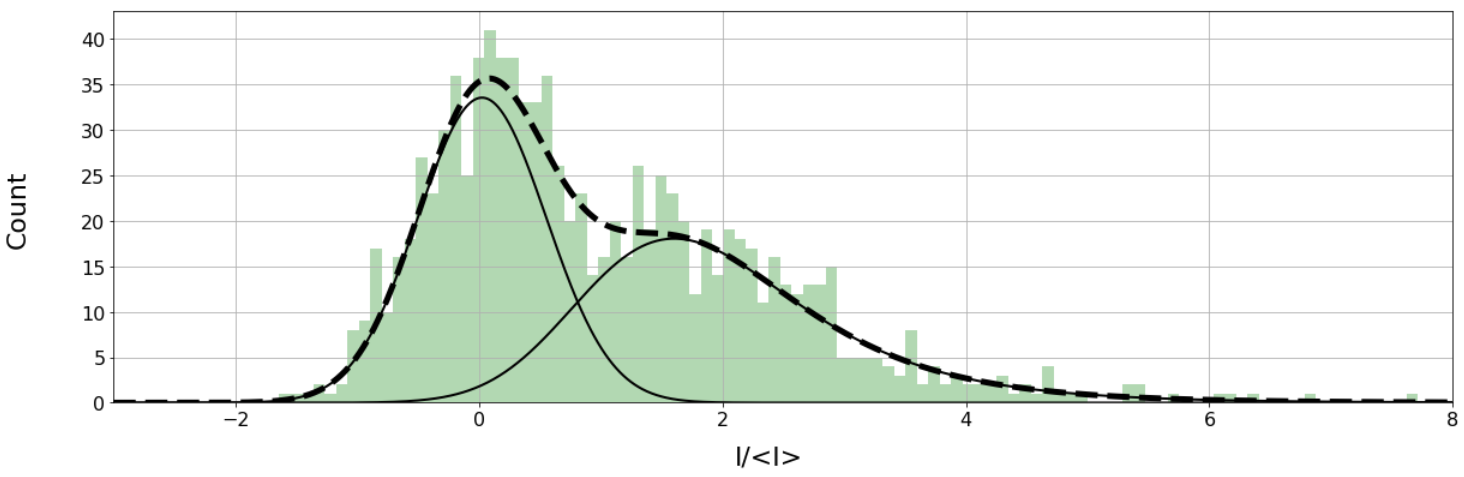}
\caption{An example histogram capturing the total flux density (intensity) within an on-pulse window during the observation of a nulling pulsar. The x-axis measures intensity, normalized by the mean. The solid black line on the left is a Gaussian component with mean and standard deviation taken from the data captured in an off-pulse window. The shape of this Gaussian should describe observation noise in the absence of any emission (suitable for null pulses in the on-pulse window or \emph{any} data in an off-pulse window). The solid line on the right is a convolution of the Gaussian function describing off-pulse noise and a lognormal distribution. Convolving permits the function to extend below zero intensity (thereby permitting some non-null pulses to have negative total flux density). The dashed line is the sum of both of the solid lines. For the BPE technique the $\mu$ and $\sigma$ of the lognormal distribution are free parameters as well as the ratio of areas under the two solid lines.}
\label{bayes_eg}
\end{figure*}

{For each observation, we define a phase window in the pulse profile where emission is expected to occur, or the \emph{on-pulse window} (e.g., the green region of Figure~\ref{J1048-5832_waterfall}). For clarity, Figure~\ref{J1048-5832_waterfall} and the other diagnostic plots for our pulsar sample (Figures~3–20) are presented together later in the paper. Its width is set manually by inspecting all available observations of a given pulsar to identify any apparent emission. The on-pulse window is chosen to encompass the specific emission component we are interested in analysing. In some pulsars, additional emission components are present at other phases. An \emph{off-pulse window} is also defined to characterize the baseline noise, representing the expected signal in the absence of pulsar emission (i.e., the null state). One might initially consider using as wide an off-pulse window as possible to improve the statistical precision of the baseline noise estimate. However, we observe that our pulsar baselines often exhibit subtle low-frequency variations with systematic trends rather than purely white noise. Extending the off-pulse window over these variations can introduce biases in the baseline estimation. Therefore, we choose a narrow off-pulse window that is always the same width as the on-pulse window and so is short compared to the timescale of the systematic drifts. This approach reduces the impact of the low-frequency variations, produces an off-pulse region consistent with white noise and allows for a more reliable baseline estimate. Additionally, a narrow off-pulse window minimizes the risk of inadvertently capturing emission from any part of the pulse profile. We adjust the flux density of the observation's pulse profile to ensure that the mean in the off-pulse window is zero. For each rotation of the pulsar, we subsequently calculate the total flux density within the on-pulse window and within the off-pulse window.

Due to the baseline fluctuations, we found that the results were sometimes sensitive to the position of the off-pulse window, whereas in ideal data all off-pulse regions would be statistically equivalent. To account for this, we performed the full analysis using two different off-pulse window locations and report the mean NF derived from these measurements. We selected two distinct locations for the off-pulse windows. The first is positioned one quarter of a pulse period later than the on-pulse window, i.e., at a pulse phase where no emission is typically expected, minimizing the likelihood of including either main pulse or interpulse emission. The second off-pulse window is placed at the beginning of the pulse profile (with the profile aligned so that the pulse peak is at bin 128, one quarter of the 512 total bins). An illustrative example is provided in Figure~\ref{J1048-5832_waterfall}.
With these preliminary analyses completed, we are now able to calculate the NFs as described below.

\subsection{Histogram Scaling Method}
\label{sec:histogram_scaling}
    
For each observation of a given pulsar, we generated one histogram each for the on-pulse and off-pulse total flux densities.  In each histogram, the sum of counts in the bins corresponds to the total number of pulsar rotations, $N$, in the observation. A scaled version of the off-pulse histogram was subtracted from the on-pulse histogram, with the scaling such that the sum of the difference counts in bins with flux density $<$ 0 (i.e. probable nulls) is zero. The calculated NF is then simply the scale factor with a fitting uncertainty of $\sqrt{n_{p}}/N$, where $n_{p} = \mathrm{NF} \times N$ is the number of null pulses.

A drawback of this method is the assumption that pulses with negative intensities are entirely caused by nulling, overlooking the possibility that pulses with low S/N can be drowned out by radiometer noise, also resulting in negative flux densities.

\subsection{Bayesian Parameter Estimation Method}
\label{sec:bayesian_parameter_estimation}


 For each observation, we generate a histogram of the summed flux density values for the on-pulse window. We expect that the probability density function (PDF) of on-pulse window flux density for \emph{null} pulses should tend to a Gaussian function with a mean close to zero.
 For the \emph{non-null} pulses the energy distribution of many pulsars is well-described by a lognormal distribution \citep{burke2012high}. When (radiometer) noise is added to the pulses, the lognormal function should be convolved by the Gaussian PDF that describes the summed flux density of the off-pulse window. See Figure~\ref{bayes_eg} for examples of curves that describe the PDFs of the null and non-null pulse populations.

Using BPE we find the combination of five parameter values which produce PDFs that maximize the likelihood that the pulsar data are described by them. Two of these variables are fixed, namely the mean and standard deviation of the Gaussian PDF that describes the on-pulse window total flux density for the population of null pulses within the observation. These are held constant as we would expect them to be close to the corresponding values for the off-window total flux density (if nulls are truly devoid of emission). There are three remaining variables: (i) the lognormal parameter \( \mu \), and (ii) the lognormal parameter \( \sigma \), both of which describe the PDF of the total flux density within the on-pulse window for the non-nulls, and (iii) the NF. The latter is the parameter in which we are ultimately interested. We use Markov chain Monte Carlo (MCMC) techniques to sample the joint posterior distribution of the model parameters. The median values and 1$\sigma$ fitting uncertainties for each parameter are then estimated from the resulting posterior distributions.

\subsection{Calculating NF Uncertainty}
\label{sec:all_unc}
Three sources of uncertainty contribute to the total error on the measured NFs. First, there is a fitting uncertainty associated with both the BPE and HS methods described above. Second, a systematic uncertainty arises from sensitivity to the choice of off-pulse window location. As detailed in Section~\ref{sec:analysis_prep}, baseline fluctuations can cause the results to vary depending on which off-pulse region is used, so we perform the analysis with two different off-pulse windows and incorporate the resulting variation into the final uncertainty. Third, a binomial uncertainty reflects the limited statistical sampling of null and emission phases within the short observation duration. We describe this component in more detail below.
\subsubsection{Statistical Sampling Uncertainty}
\label{sec:binom_unc}
Because our goal is to estimate the underlying, intrinsic NF of each pulsar (as a function of time), not just the value measured in one observation, quantifying the associated statistical uncertainty is essential. In nulling pulsars, observations often capture consecutive trains of nulls and of non-nulls. Finite observation can merely sample this phenomenon and so the NF becomes subject to binomial statistics; calculating the binomial uncertainty is crucial for precise analysis. In this context, the number of \emph{trials} is considered to be the number of times the pulsar would typically switch between a train of emission and one of nulling over the course of an observation.
To estimate the number of trials for each observation, we analysed the length of consecutive trains of nulls and non-nulls, commonly referred to as the \emph{nulling duration} and \emph{burst duration}, respectively. Given the inherent uncertainty in distinguishing null pulses from non-null pulses (accurately calculating the NF is a significant part of this study after all), we adopted a simplified method to classify each pulse for this stage of the analysis. For each pulsar, we set a threshold flux density. If the flux density of a pulse fell below this threshold, it was categorized as a null; otherwise, it was considered a non-null. This threshold was set to be the mean, plus one standard deviation of the flux density in the off-pulse window for the pulsar, ensuring that only pulses with somewhat significant emission were classified as non-nulls. After applying this classification across all observations for a given pulsar, we measured the lengths of consecutive nulling and bursting intervals and computed their mean values. These typical durations then allowed us to estimate the number of switching events (i.e., trials) within each observation by dividing the total observing time by the mean train length.}
These trial counts were then used to calculate the binomial statistics necessary for analyzing the uncertainty in the NF.

We combined the fitting and binomial uncertainties in quadrature for each NF measurement, then calculated the final NF as the weighted mean of the two values from the different off-pulse windows. The uncertainty on this weighted mean accounts for both the individual measurement errors and their difference, thus incorporating fitting, statistical, and systematic uncertainties.

\subsection{S/N Estimates}
\label{sec:sn_est}
We approximate the typical single-pulse S/N for each observation to serve as a useful diagnostic metric. To do so we first ensure that we calculate it using only genuine pulses of emission. We achieve this by selecting the most prominent non-null pulses, specifically those with S/N values (measured as single pulse peak flux density divided by the standard deviation of flux density in the off-pulse window) above the 95th percentile for the observation. We then take the median of these values to calculate a representative single-pulse S/N proxy for each observation (e.g., the upper panel of Figure~\ref{J1048-5832_nf_evolution}). Using the median ensures that this estimate is not strongly affected by occasional unusually bright pulses. This S/N proxy is intended solely as a qualitative indicator of data quality and primarily reflects the brightest pulse components rather than the full pulse-energy distribution.

\section{Testing the NF techniques with simulated data}
\label{sec:simulations}
\par{To compare the two NF calculation techniques described in Section~\ref{sec:histogram_scaling} and Section~\ref{sec:bayesian_parameter_estimation}, we simulate observations of a nulling pulsar and try to recover the NF. To test the two techniques under different conditions, we simulate pulsar observations with three different fundamental NF values of 0.2, 0.5, and 0.8, and four observation scenarios for each NF value:
\begin{enumerate}[label=\roman*.]
\item 200 single pulses with S/N $\sim$ 2.0
\item 1000 single pulses with S/N $\sim$ 2.0
\item 200 single pulses with S/N $\sim$ 10.0
\item 1000 single pulses with S/N $\sim$ 10.0
\end{enumerate}
In general, calculating the NF becomes more accurate as the number of single pulses in the observation grows and when the S/N of the single pulses is high. In these simulations, the fundamental NF corresponds to the probability that any given single pulse will be a null (i.e. 20\%, 50\%, or 80\%). As the observation length increases, the observed NF tends to converge toward the fundamental NF, but will deviate from it when the observation length is finite. Specifically, the standard deviation of the measured NF scales as $\sqrt{N}$, where $N$ is the number of rotations in the simulated observation. The simulated 512-phase-bin pulse profile contains a single Gaussian component whose mean amplitude is set by the chosen S/N, while the amplitude of each individual pulse is drawn from a lognormal distribution.

\subsection{Results from the Two NF Calculation Methods}

Figure~\ref{comparison} shows the results from calculating the NF 1,000 times using both the HS and BPE techniques across three fundamental NF values (0.2, 0.5, and 0.8) and the four observation scenarios described in Section~\ref{sec:simulations}. In all twelve scenarios, the BPE method recovers the true NF more reliably than the HS method. Beyond its overall accuracy, the BPE technique represents a conceptually important improvement over existing NF-estimation methods. Because it explicitly models the on-pulse energies as a mixture of (i) a Gaussian noise distribution describing the nulls and (ii) a lognormal distribution describing the non-nulls, it naturally incorporates the well-established result that the energies of non-null pulses in many pulsars follow a lognormal form. This allows the method to solve simultaneously for the underlying lognormal parameters and the NF.

The performance of the BPE method also follows the expected trends with S/N and observation length. For simulations with single-pulse S/N $\approx 10$, the recovered NFs cluster extremely tightly around the true values: in all three NF regimes, most realisations differ from the input NF by $\lesssim 0.01$. Increasing the number of simulated pulses from 200 to 1000 further narrows the spread in recovered NF, as the posterior distribution becomes better constrained with more data. At lower S/N ($\approx 2$), the scatter increases but the BPE method remains stable and consistently outperforms the HS technique.
We emphasise, however, that these simulations represent an idealised scenario in which the on-pulse energies follow a perfect lognormal distribution, the off-pulse noise is strictly Gaussian, and the baseline is flat. Real pulsar observations inevitably deviate from these assumptions, meaning that the relative performance of the two techniques on actual data may differ. Even so, the simulations provide a useful benchmark for understanding how each method behaves in the absence of observational systematics.

Having assessed the efficacy of both techniques on simulated pulsar data, we apply them to PMPS observations for selected pulsars that either show full nulls or exhibit a nulling component.

\begin{figure*}
\centering
\includegraphics[width = 1.0\textwidth]{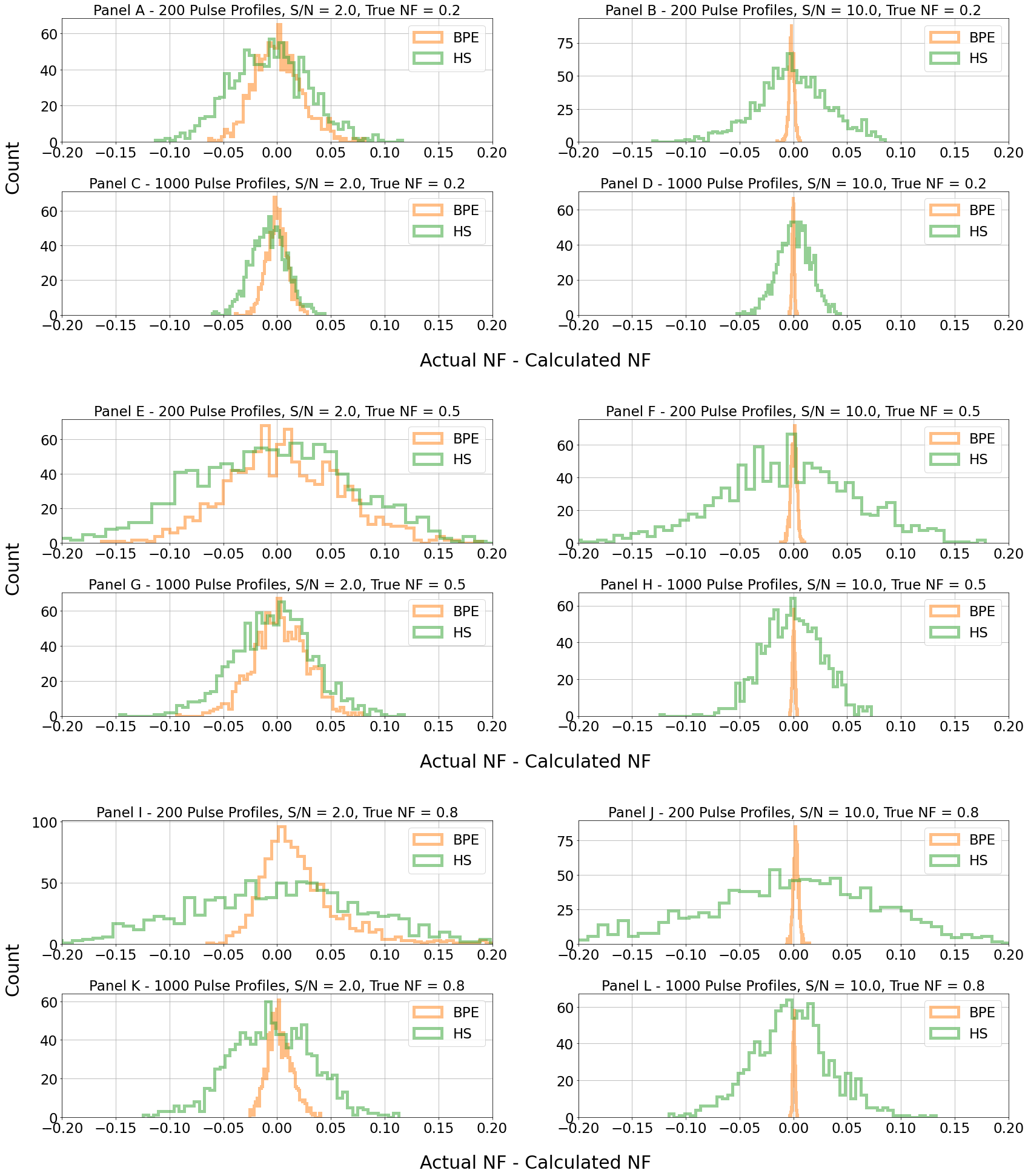}

\caption{The difference between actual NF and the calculated NF for simulated nulling-pulsar observations (described in Section~\ref{sec:simulations}). We compare the BPE and HS NF-calculation methods across three true NF values (0.2, 0.5, and 0.8) and four scenarios per NF: 200 pulse profiles with S/N = 2.0, 200 pulse profiles with S/N = 10.0, 1000 pulse profiles with S/N = 2.0, and 1000 pulse profiles with S/N = 10.0. In each scenario the NF was computed 1,000 times, with the data being generated anew for each iteration.}
\label{comparison}
\end{figure*}

\section{Pulsars Analyzed}
\label{sec:pulsars}
As well as the information below, pulse periods, flux densities, surface magnetic fields and spindown ages for the analyzed pulsars can be found in Table~\ref{tab:pulsar_table}.
\subsection{PSR~J1048$-$5832}
\label{PSRJ1048}
\par{PSR~J1048$-$5832 was discovered by the Parkes high-frequency survey \citep{10.1093/mnras/255.3.401} and has subsequently also been observed in the x-ray \citep{Gonzalez_2006} and gamma-ray \citep{2010ApJ...713..154A} domains. While initial inspection suggests it could be a nulling pulsar, \citet{Yan_2019} show that its pulse energy remains above zero even during apparent null states. The authors consequently designate it as a mode-changing pulsar rather than a nulling pulsar. However, the pulse profile consists of two components and the trailing-edge component appears to null, even if the overall profile does not (see Figure~\ref{J1048-5832_waterfall}}). PSR~J1048$-$5832 has also been seen to exhibit glitching behaviour \citep[e.g.,][]{10.1046/j.1365-8711.2000.03713.x}. We have analyzed 97 observations of this pulsar.

\subsection{PSR~J1114$-$6100}
\par{This pulsar has previously been seen to display short-duration nulls and is know to have an extremely high rotation measure; the fourth highest of any known pulsar despite being located far from the Galactic centre \citep{10.1093/mnras/stab095}. The pulse profile resembles a singular Gaussian component (see Figure~\ref{J1114-6100_waterfall}). We have analyzed 31 observations of PSR~J1114$-$6100.}


\subsection{PSR~J1453$-$6413}
\par{PSR~J1453$-$6413 is a middle-aged pulsar that has been seen to glitch \citep{2019MNRAS.489.3810P, 2023RAA....23j5014L} and was categorized as a nulling pulsar in the HTRU Survey. The pulse profiles we observe have a singular Gaussian-like component, albeit displaying slight temporal broadening (see Figure~\ref{J1453-6413_waterfall}}). We have analyzed 49 observations of this pulsar.
    
\subsection{PSR~J1502$-$5653}
\par{PSR~J1502$-$5653 was characterized as a nulling pulsar in the HTRU Survey and the NF was previously calculated to be 93\% by \citet{2007MNRAS.377.1383W}. PSR~J1502$-$5653 displays tens to hundreds of consecutive detectable pulses interspersed with intervals of thousands of null pulses. 
The pulse profile has a single sharp and narrow Gaussian-like component, covering only a small fraction of the pulse period (see Figure~\ref{J1502-5653_waterfall}). We have analyzed 31 observations of this pulsar. One observation appears to capture no pulses of emission and so has an NF close to 1.0 (see Figure~\ref{J1502-5653_nf_evolution}). Because of how it is calculated (see Section~\ref{sec:sn_est}), the S/N proxy value will be especially low for this observation (as there is no signal).}
    
\subsection{PSR~J1559$-$5545}
\par{This pulsar was characterized as a nulling pulsar in the HTRU Survey. The pulse profile appears to be comprised of two overlapping emission components. As a result the pulse profile is a sharp and narrow Gaussian-like component with a leading shoulder (see Figure~\ref{J1559-5545_waterfall}). The two components seem to null quasi-independently, and both fall within our on-pulse window. 
We have analyzed 29 observations of this pulsar.
    
\subsection{PSR~J1745$-$3040}
\label{psrj1745}
The pulse profile of PSR~J1745$-$3040 consists of a small leading component and a more complex trailing component. We have analyzed the nulling behaviour of both components separately (see Figures~\ref{J1745-3040_major_waterfall} and \ref{J1745-3040_minor_waterfall}).
This pulsar was characterized as a nulling pulsar in the HTRU Survey. However, \citet{2021MNRAS.502.1253J} state that nulling is not evident in PSR~J1745$-$3040. The authors do note that various parts of the profile seem to switch off at various times and the flux in the leading component is often anti-correlated with the flux of the trailing components.
We have analyzed 35 observations of this pulsar. Multiple quasi-independent nulling components are encompassed when our on-pulse window is placed around the trailing pulse-profile feature.

\subsection{PSR~J1825$-$0935}
\par{PSR~J1825–0935 (B1822–09) is a well-known mode-changing pulsar with three main components in its pulse profile: a main pulse, a precursor component leading the main pulse and an interpulse (see Figure~\ref{J1825-0935_waterfall}). The pulsar displays mode changing and sub-pulse modulation with strong anti-correlation between the interpulse and precursor components \citep[][]{2010MNRAS.404...30B}. PSR~J1825$-$0935 has two emission modes: a \emph{quiet} mode where the precursor is absent (or drops to a very low level) and the interpulse is present, and a \emph{bright} mode where this situation is reversed. We have focused on the nulling nature of the precursor component which we place within the on-pulse window. The average time between the mode change is a few hundred pulse periods \citep{1982A&A...109..279F, 2012MNRAS.427..180L}; in our observations the null sequences are of a comparable length to that of our observations. Because of this, some observations catch the pulsar entirely (or almost entirely) in either the quiet or bright mode, resulting in the apparently bimodal split of NF seen in Figure~\ref{J1825-0935_nf_evolution}. We have analyzed 65 observations of this pulsar.}

\subsection{PSR~J1847$-$0402}
PSR~J1847$-$0402 was discovered in a systematic search at low Galactic latitudes with the Lovell Telescope at the Jodrell Bank Observatory \citep{1970Natur.227.1123D}. The pulsar has been seen to glitch \citep{10.1093/mnras/stab3336, 10.1093/mnras/stab2678} and was categorized as a nulling pulsar in the HTRU Survey. The pulse profile of PSR~J1847$-$0402 consists of two overlapping Gaussian-like components, which appear to null quasi-independently, and therefore both fall within our on-pulse window (see Figure~\ref{J1847-0402_waterfall}). We have analyzed 41 observations of this pulsar.

\section{Results}
\label{sec:results}
The NF for each pulsar in our study is measured across multiple observations spanning 8 to 10 years.
The NF results are represented in Figures~\ref{J1048-5832_nf_evolution}, \ref{J1114-6100_nf_evolution}, \ref{J1453-6413_nf_evolution}, \ref{J1502-5653_nf_evolution}, \ref{J1559-5545_nf_evolution}, \ref{J1745-3040_major_nf_evolution}, \ref{J1745-3040_minor_nf_evolution}, \ref{J1825-0935_nf_evolution} and \ref{J1847-0402_nf_evolution}, and the mean and standard deviation of the measured NFs are summarised in Table~\ref{tab:meannf_table}.
We carried out a regression analysis on each pulsar dataset to determine the best-fit gradient and intercept parameters for the relationship between NF and time. These parameters were calculated along with their associated uncertainties. The uncertainty on each fitted slope $a$ was taken from the diagonal of the parameter covariance matrix (yielding $\sigma_a$), and the significance of a trend is calculated as $|a|/\sigma_a$ in Table~\ref{tab:pulsar_table}, which shows the NF gradients.
Although most pulsars analysed show no significant trend in their NF over time, three pulsars show evidence for non-zero gradients in NF at the $>3\sigma$ level in at least one of the two measurement methods: PSRs~J1048$-$3832, J1745$-$3040, and J1825$-$0935 (Table ~\ref{tab:pulsar_table}). PSR~J1048$-$3832 is the only pulsar that shows a $>3\sigma$ gradient in both NF estimation methods, with significances of $4.3\sigma$ (BPE) and $4.8\sigma$ (HS). For PSR~J1825$-$0935, the inferred gradient is significant at the $>3\sigma$ level in the BPE method ($3.1\sigma$), but not in the HS method ($1.0\sigma$). In the case of PSR~J1745$-$3040, the significance depends on the pulse-profile component under analysis: the main component shows significances of $3.0\sigma$ (BPE) and $0.5\sigma$ (HS), while the leading minor component shows significances of $0.7\sigma$ (BPE) and $4.3\sigma$ (HS). We discuss these results in more detail in the next section.

\begin{deluxetable*}{ccccccccc}
\tablecaption{Pulsar properties, NF gradients, and their standard deviations from zero. Pulse periods, mean flux densities, inferred surface magnetic fields and spindown age are taken from the ATNF Catalog \citep{Manchester2005}.\label{tab:pulsar_table}}
\tablehead{
\colhead{\shortstack{Pulsar\\Name}} &
\colhead{\shortstack{NF Gradient\\BPE\\($10^{-3}$ NF/yr)}} &
\colhead{\shortstack{SDs\\from\\Zero}} &
\colhead{\shortstack{NF Gradient\\HS\\($10^{-3}$ NF/yr)}} &
\colhead{\shortstack{SDs\\from\\Zero}} &
\colhead{\shortstack{Pulse\\Period\\(ms)}} &
\colhead{\shortstack{Flux\\Density\\(mJy)}} &
\colhead{\shortstack{Surface B\\($\times10^{12}$ G)}} &
\colhead{\shortstack{Spindown\\Age\\(yr)}}
}
\startdata
J1048$-$5832 & $-3.0 \pm 0.7$   & 4.3 & $-2.7 \pm 0.6$   & 4.8 & 123.7 & 9.1  & 3.5 & $2.05 \times 10^4$ \\
J1114$-$6100 & $-1.5 \pm 4.9$   & 0.3 & $ 2.0 \pm 3.5$   & 0.6 & 880.9 & 5.4  & 6.4 & $3.03 \times 10^5$ \\
J1453$-$6413 & $-1.2 \pm 1.9$   & 0.6 & $-0.3 \pm 1.9$   & 0.1 & 179.5 & 18.0 & 0.7 & $1.04 \times 10^6$ \\
J1502$-$5653 & $ 3.1 \pm 3.0$   & 1.0 & $-3.6 \pm 2.3$   & 1.6 & 535.5 & 0.4  & 1.0 & $4.64 \times 10^6$ \\
J1559$-$5545 & $-20.1 \pm 10.3$ & 2.0 & $ 9.2 \pm 4.7$   & 2.0 & 957.2 & 0.7  & 4.4 & $7.62 \times 10^5$ \\
J1745$-$3040 (Main)  & $ 9.4 \pm 3.1$ & 3.0 & $-1.7 \pm 3.3$ & 0.5 & 367.4 & 21.0 & 2.0 & $5.46 \times 10^5$ \\
J1745$-$3040 (Minor) & $ 2.6 \pm 3.9$ & 0.7 & $12.2 \pm 2.9$ & 4.3 & ---   & ---  & --- & --- \\
J1825$-$0935 & $-11.6 \pm 3.7$  & 3.1 & $-3.6 \pm 3.8$   & 1.0 & 769.0 & 10.0 & 6.4 & $2.33 \times 10^5$ \\
J1847$-$0402 & $-0.2 \pm 2.1$   & 0.1 & $-1.2 \pm 1.9$   & 0.6 & 597.8 & 4.1  & 5.6 & $1.83 \times 10^5$ \\
\enddata
\end{deluxetable*}


\begin{deluxetable*}{lcccc}
\tablecaption{Mean and standard deviation of the calculated NFs for each pulsar.\label{tab:meannf_table}}
\tablehead{
\colhead{Pulsar Name} & \colhead{Mean of NF for BPE} & \colhead{SD of NF for BPE} &
\colhead{Mean of NF for HS}  & \colhead{SD of NF for HS}
}
\startdata
J1048$-$5832            & 0.04 & 0.04 & 0.06 & 0.04 \\
J1114$-$6100            & 0.11 & 0.09 & 0.19 & 0.10 \\
J1453$-$6413            & 0.06 & 0.02 & 0.04 & 0.02 \\
J1502$-$5653            & 0.93 & 0.04 & 0.92 & 0.04 \\
J1559$-$5545            & 0.28 & 0.16 & 0.47 & 0.15 \\
J1745$-$3040 (Main)     & 0.46 & 0.08 & 0.53 & 0.13 \\
J1745$-$3040 (Minor)    & 0.37 & 0.12 & 0.42 & 0.10 \\
J1825$-$0935            & 0.60 & 0.36 & 0.62 & 0.34 \\
J1847$-$0402            & 0.03 & 0.03 & 0.04 & 0.05 \\
\enddata
\end{deluxetable*}

\vspace*{\baselineskip}

\section{Discussion}
\label{sec:discussion}
We have computed NFs for eight pulsars across multiple observations spanning close to a decade. 
The primary objective of our analysis was to chronicle, for the first time, the evolutionary trajectory of nulling phenomena within pulsars over the course of many years.

For the majority of pulsars in our sample, the inferred NF gradients are consistent with zero within uncertainties, indicating no significant long-term evolution in their nulling behaviour.

PSR~J1048$-$3832 is the only pulsar in our sample for which the fitted NF gradient is inconsistent with zero at greater than the $3\sigma$ level using both NF estimation methods, with significances of $4.3\sigma$ (BPE) and $4.8\sigma$ (HS). In both cases, the inferred gradients are negative, with values of $-3.0 \pm 0.7$ and $-2.7 \pm 0.6 \times 10^{-3}\,\mathrm{NF/yr}$, respectively (Figure~\ref{J1048-5832_nf_evolution}). The pulse profile of PSR~J1048$-$3832 consists of two components (Figure~\ref{J1048-5832_waterfall}). As discussed in Section~\ref{PSRJ1048}, the trailing-edge component appears to null even if the overall pulse profile does not. For this component, an NF gradient of order $-3 \times 10^{-3}\,\mathrm{NF/yr}$ implies that, if the trend reflects genuine long-term evolution and persists at a similar rate, the incidence of nulling would decrease substantially on decadal timescales, potentially approaching a zero NF within a couple of decades. While such linear extrapolation should be treated with caution, this illustrates that the measured gradient corresponds to potentially significant evolution in the emission behaviour of this pulsar.

For PSR~J1825$-$0935, the fitted NF gradient is inconsistent with zero at the $>3\sigma$ level when estimated using the BPE method ($3.1\sigma$), but is consistent with zero when using the HS method ($1.0\sigma$). The BPE-derived gradient corresponds to a decrease in NF of $-11.6 \pm 3.7 \times 10^{-3}\,\mathrm{NF/yr}$ (Figure~\ref{J1825-0935_nf_evolution}). PSR~J1825$-$0935 is known to exhibit two emission modes: a quiet mode in which the precursor component is absent or strongly suppressed while the interpulse is present, and a bright mode in which the precursor is prominent and the interpulse is weak or absent. Our analysis focuses on the nulling behaviour of the precursor component. If the BPE-derived trend reflects genuine long-term evolution, the observed decrease in NF could arise from an increasing fraction of time spent in the bright mode, a reduction in the degree of nulling within one or both emission modes, or a combination of both effects. Although the two NF estimation methods yield different gradient significances, most individual NF measurements are in reasonable agreement between methods (Figure~\ref{J1825-0935_nf_evolution}). The discrepancy in inferred gradients likely reflects the sensitivity of the regression to modest differences in individual measurements or their uncertainties, such that small changes in a subset of epochs can lead to appreciable differences in the fitted long-term trend.

For PSR~J1745$-$3040 we tracked the NF evolution separately for the main pulse component (Figure~\ref{J1745-3040_major_nf_evolution}) and a minor pulse component (Figure~\ref{J1745-3040_minor_nf_evolution}). The two NF-calculation methods yield different inferences regarding the fitted NF gradients. For the main component, the BPE method returns a positive gradient of $9.4 \pm 3.1 \times 10^{-3}\,\mathrm{NF/yr}$, inconsistent with zero at the $3.0\sigma$ level, whereas the HS method gives $-1.7 \pm 3.3 \times 10^{-3}\,\mathrm{NF/yr}$, consistent with zero ($0.5\sigma$). One contributor to this discrepancy is a limitation of the HS technique.
For several observations of PSR~J1745$-$3040, the histogram of summed on-pulse flux densities for the main component, shows a distorted negative tail, with an excess of counts forming an off-centre, non-Gaussian feature relative to the off-pulse histogram.
Because the HS method estimates the NF solely from the negative-flux values, such excesses can bias individual NF measurements.
In contrast, the BPE technique models the full distribution of on-pulse flux densities rather than relying only on the negative tail, and is therefore affected differently by such distortions.
These biases can then propagate into the HS-derived gradient, leading to differences between the long-term trends inferred by the two NF-calculation methods.
For the minor component, the BPE method yields a gradient of $2.6 \pm 3.9 \times 10^{-3}\,\mathrm{NF/yr}$, consistent with zero ($0.7\sigma$), while the HS method recovers a positive gradient of $12.2 \pm 2.9 \times 10^{-3}\,\mathrm{NF/yr}$, inconsistent with zero at the $4.3\sigma$ level.
In practice, the NF values recovered by the two methods are similar for the majority of observations for the minor component of PSR~J1745$-$3040. However, modest differences in a small number of epochs are sufficient to change the fitted gradient, and because the formal uncertainty on the slope is relatively small, these differences can lead to a substantial change in the inferred significance.

While some analysis methods yield formally significant non-zero NF gradients, as seen in specific cases discussed above, the inferred slopes can be sensitive to reasonable methodological choices. This sensitivity reflects the fact that modest differences in a small number of NF measurements can propagate into the fitted gradient, suggesting that the formal uncertainties may underestimate the true error budget and that the evidence for non-zero gradients should be regarded as tentative in some cases.
In practice, this methodological sensitivity often arises from observational systematics that affect individual NF measurements rather than from the fitting procedure alone.

One such observational systematic is the stability of the pulse-profile baseline. Both NF-calculation techniques we have employed in this work are sensitive to the integrity of the pulse-profile baseline within each observation. 
Noise or other confounding factors can compromise the baseline, introducing inaccuracies and inconsistencies in the inferred NFs.
We tested applying additional baseline-flattening steps beyond the standard processing, but found that this distorted the noise statistics: the mean and width of the on-pulse null distribution no longer matched those of the off-pulse noise, even though true null pulses should resemble the off-pulse noise distribution.
This indicated that further baseline removal risked biasing the flux-density histograms rather than improving them.

In the specific case of PSR~J1745$-$3040, we also found that a subset of observations exhibited pronounced distortions in the pulse-profile baseline. Such epochs occurred preferentially at later times and tended to yield systematically higher NF values than the rest of the data, consistent with baseline distortions beneath the pulse profile artificially inflating the inferred NF.
Because these observations could therefore exert a disproportionate influence on the fitted NF gradients, they were excluded from the primary analysis. Appendix~A discusses these baseline distortions in detail, and Figure~\ref{J1745-3040_removed} shows all pulse profiles for PSR~J1745$-$3040, including those excluded from the analysis.

In addition to baseline-related effects, distortions of the on-pulse noise distribution can bias NF measurements in a systematic way, as seen in the NF calculations for the main pulse component of PSR~J1745$-$3040. For the HS method, any distortion of the on-pulse noise distribution, whether producing an excess or a deficit of negative-flux values, will bias the inferred NF because the method depends solely on the negative tail of the total-flux-density histogram; although the BPE method can also be affected by such distortions, it is far less sensitive since it uses the full on-pulse flux-density distribution.
Furthermore, both techniques operate under the presumption that some fraction of the pulses within the on-window correspond to null states. Should these pulses merely exhibit low emission rather than true nulls, the resulting NF calculations will be inaccurate; nulls within the on-pulse window must resemble the noise observed in the off-pulse window for more accurate NF results. For example, if the section of the pulse profile under analysis in the on-pulse window consists of multiple overlapping emission components (PSR~J1048$-$5832, PSR~J1559$-$5545, the trailing component of PSR~J1745$-$3040 and PSR~J1847$-$0402, see Figures~\ref{J1048-5832_waterfall}, \ref{J1559-5545_waterfall}, \ref{J1745-3040_major_waterfall} and \ref{J1847-0402_waterfall}), and only parts of the profile null rather than the feature as a whole, then the NF calculation methods we have employed will not be optimal.
A related point is that if we use our NF-calculation method on a non-nulling pulsar using low S/N observations, the result will be a non-zero NF due to the nature of the techniques.

A further consideration is that the BPE method works on the assumption that the energy distribution of a pulsar's non-null pulses conforms to a lognormal distribution. Should this assumption not hold true, it would adversely affect the reliability of the NF calculations. Despite the potential inaccuracies in determining NF in particular pulsars, however, the analyses carried out remain valuable for tracking changes in emission behaviour over years and decades.

Although the interpretation of long-term NF trends is necessarily conditional on their robustness, it is nonetheless informative to consider the physical processes that could give rise to such evolution. If the NF trends identified here do reflect real physical evolution, a gradual change in a component’s nulling fraction could plausibly arise from either slow evolution of the magnetospheric emission processes or from a gradual geometric drift of the line of sight across the emission beam.

Longer observations of nulling pulsars with sensitive instruments will provide more accurate and precise values for NF. It has been at least 14 years since the pulsar data analysed in this paper were recorded. Because of this, even a single high S/N contemporary observation would provide useful information about NF gradients and, consequently, about the emission processes in pulsars.

During the course of this study, we became aware of an alternative NF calculation method introduced by \citet{2018ApJ...855...14K}. This method introduces another Bayesian algorithm based on Gaussian mixture models. In future work, applying this Gaussian mixture method could be a valuable way to further cross-check our results.

\section{Conclusions}
\label{sec:conclusions}

Using two alternative techniques, we have computed the NF for eight pulsars over approximately a decade of observations. For most pulsars in our sample, the inferred NF gradients are consistent with zero. Statistically significant gradients are recovered in a small number of cases, most notably PSR~J1048$-$3832, while weaker and method-dependent trends are seen in PSRs~J1745$-$3040 and J1825$-$0935.

As discussed above, the inferred gradients are sensitive to the choice of NF-estimation method, baseline stability, and modest differences in individual measurements. These sensitivities imply that none of the gradients reported here should be regarded as definitive measurements of long-term evolution. Rather, they indicate that changes in nulling behaviour on decadal timescales may be present in some pulsars and are potentially measurable with sufficiently high-quality data.

If real, such changes could reflect slow evolution of the pulsar magnetosphere or gradual geometric drift of the emission beam. Future high-sensitivity, multi-epoch observations will be essential for establishing whether the trends identified here persist and for determining their physical origin.



\begin{figure*}
\centering
\includegraphics[width = 1.0\textwidth]{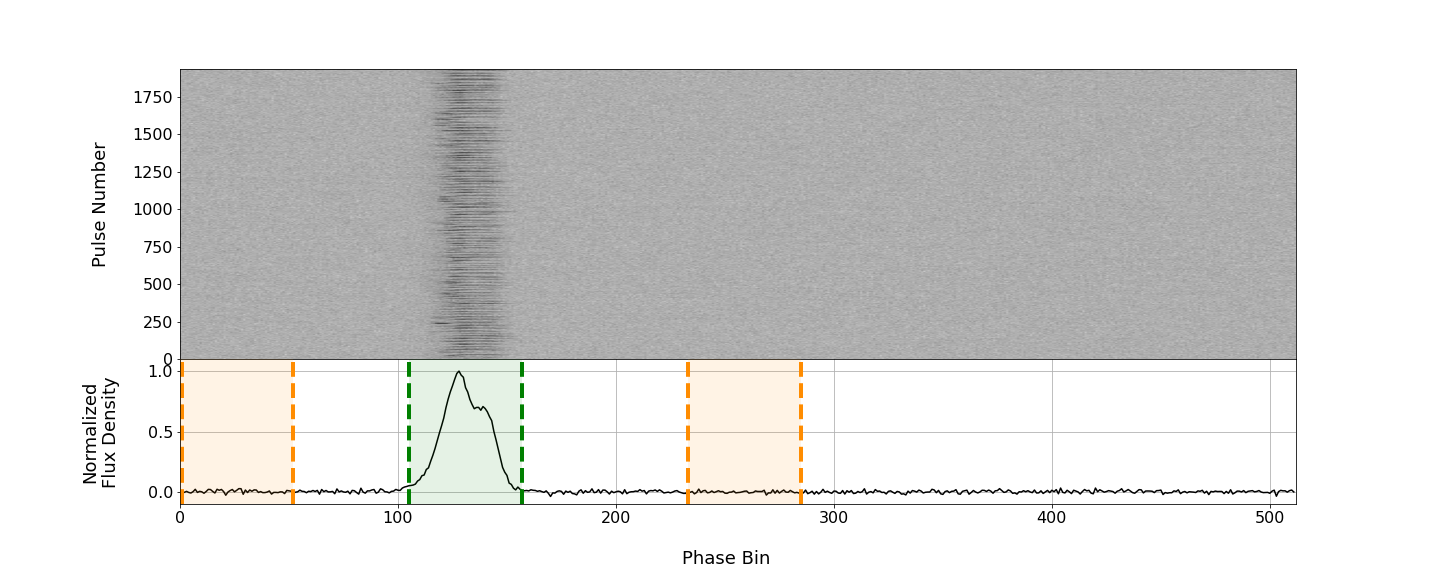}

\caption{Observation of PSR~J1048$-$5832 made on MJD 51880. The upper panel shows the intensity of the pulsar signal as a function of pulse phase on the horizontal axis and time, or pulse number, on the vertical axis. The lower panel shows the normalized average of all the single pulses that feature in the upper panel. The central highlighted region marks the on-pulse window, while the outer highlighted regions indicate the off-pulse windows. The pulse profile consists of two components and the trailing-edge component appears to null, even if the overall profile does not. The calculated NF for this observation was $0.03 \pm 0.01$
 and $0.06 \pm 0.01$ for the BPE and HS methods respectively.}
\label{J1048-5832_waterfall}
\end{figure*}

\begin{figure*}
\centering
\includegraphics[width = 1.0\textwidth]{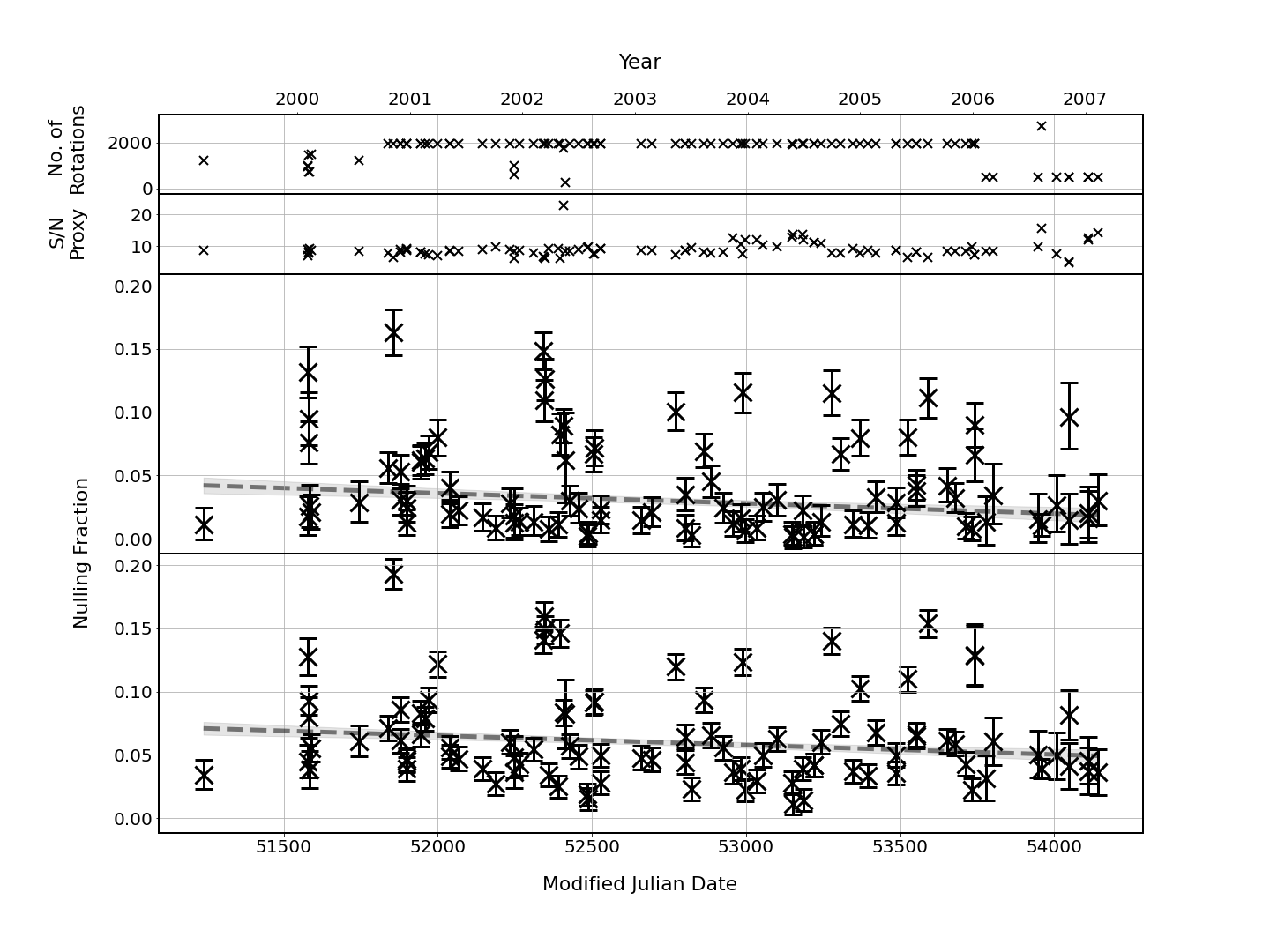}

\caption{Evolution of NF for PSR~J1048$-$5832 as a function of time. Top panel: The number of pulsar rotations recorded during each observation. Second panel: An esimation of the S/N of each observation (see Section~\ref{sec:sn_est}). Third Panel: NFs as calculated by the BPE technique. Error bars show the total $1\sigma$ uncertainty on the NF measurements, incorporating fitting uncertainty, statistical (binomial) sampling uncertainty, and systematic uncertainty associated with the choice of off-pulse window (Section~\ref{sec:all_unc}). The dashed regression line represents the best-fit linear model for the data points and weighted by their uncertainties. The gray uncertainty band around the regression line represents the 95\% confidence interval for the predicted values, accounting for the uncertainty in both the slope and the intercept, illustrating the range of plausible predictions, given the model's uncertainty and data variability. Bottom panel: As the third panel but for the HS technique.}
\label{J1048-5832_nf_evolution}
\end{figure*}


\begin{figure*}
\centering
\includegraphics[width = 1.0\textwidth]{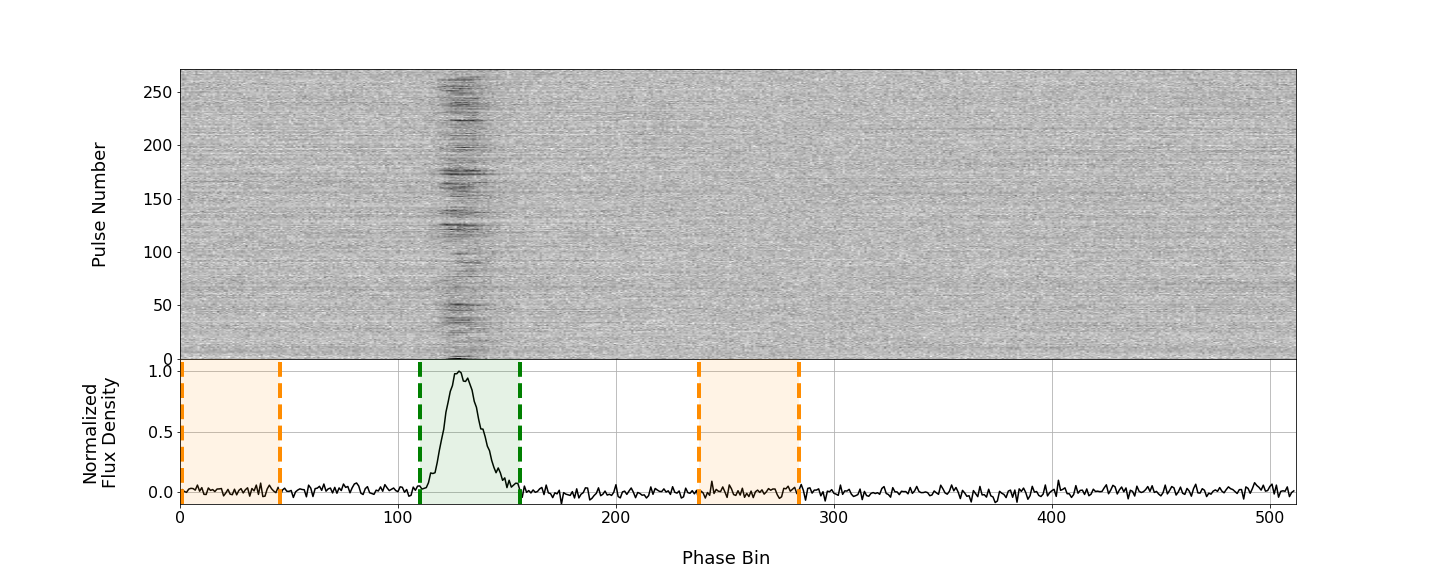}

\caption{Observation of PSR~J1114$-$6100 made on MJD 52265. The pulse profile resembles a singular Gaussian component. The calculated NF for this observation was $0.14 \pm 0.08$ and $0.21 \pm 0.03$ for the BPE and HS methods respectively. Same format as Figure~\ref{J1048-5832_waterfall} otherwise.}
\label{J1114-6100_waterfall}
\end{figure*}

\begin{figure*}
\centering
\includegraphics[width = 1.0\textwidth]{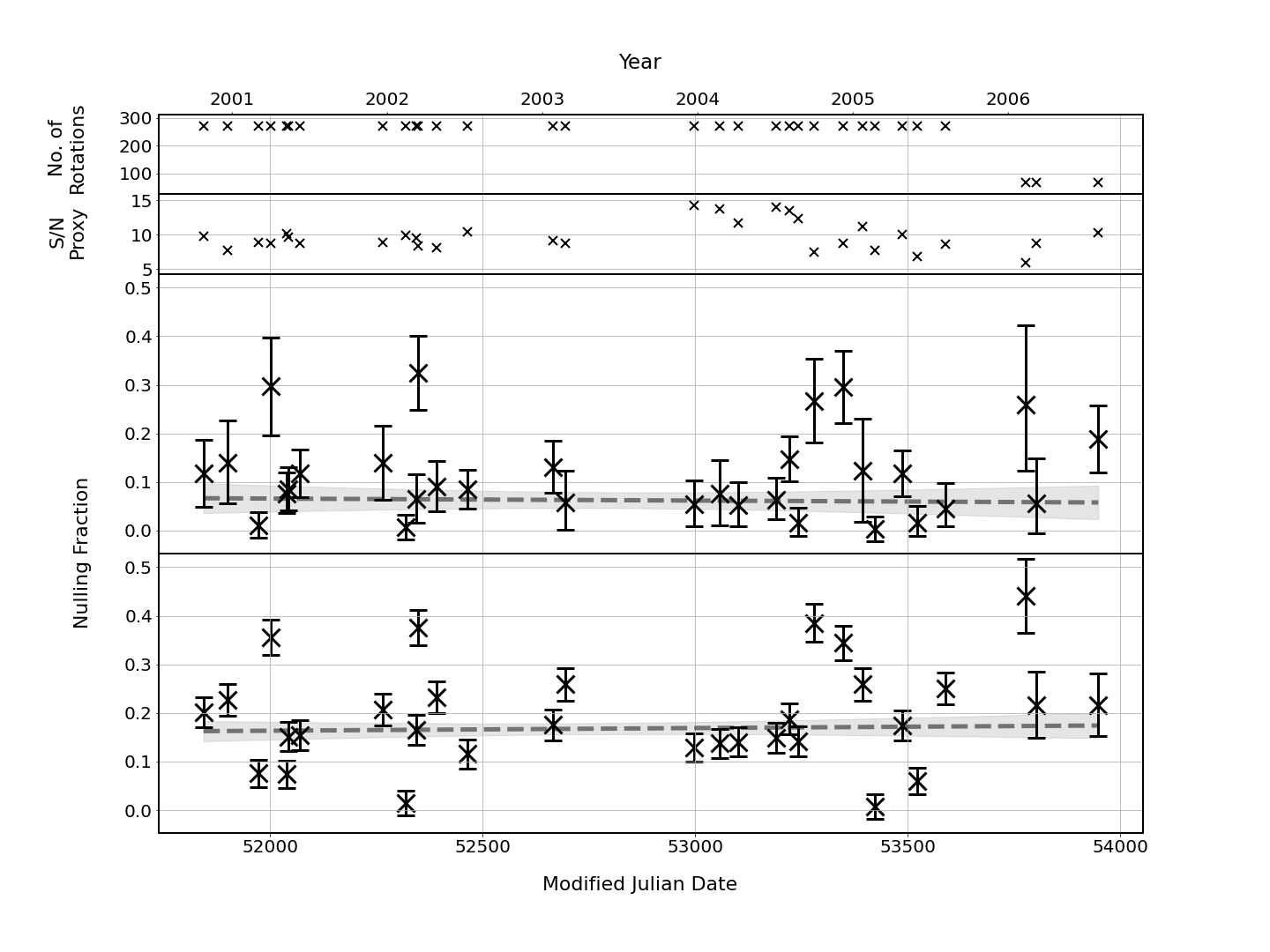}

\caption{Evolution of NF for PSR~J1114$-$6100. Same format as Figure~\ref{J1048-5832_nf_evolution} otherwise.}
\label{J1114-6100_nf_evolution}
\end{figure*}







\begin{figure*}
\centering
\includegraphics[width = 1.0\textwidth]{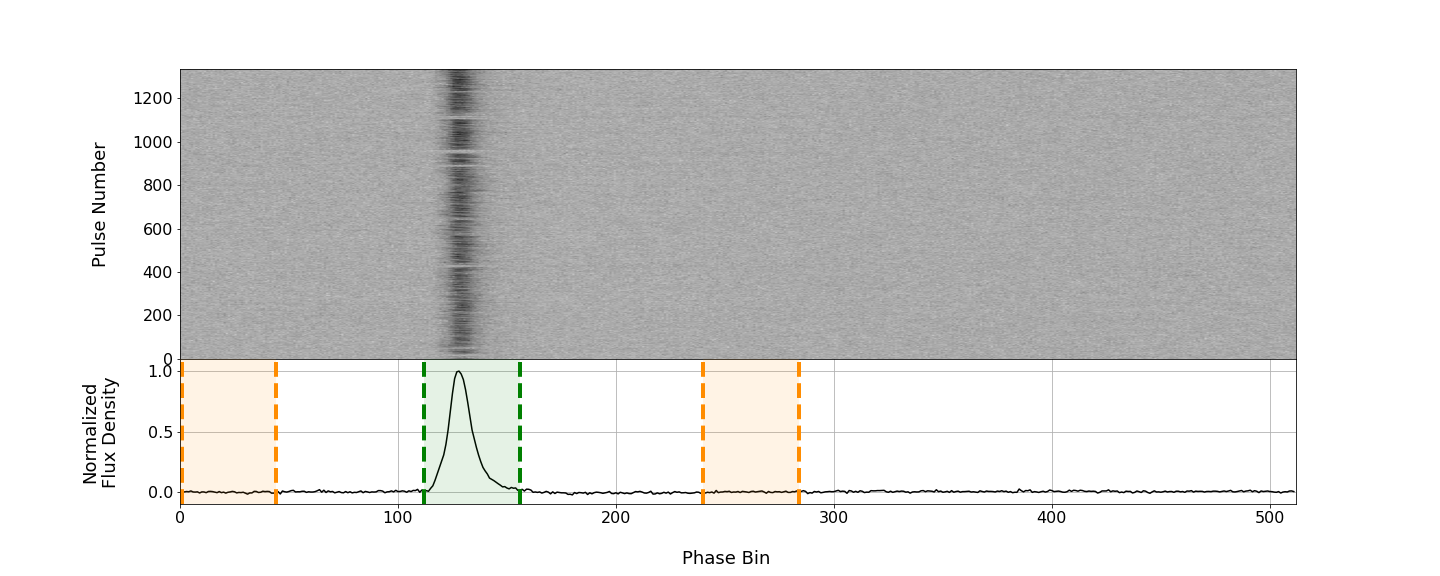}

\caption{Observation of PSR~J1453$-$6413 made on MJD 53394. The pulse profiles we observe have a singular Gaussian-like component with slight temporal broadening. The calculated NF for this observation was $0.06 \pm 0.02$ and $0.04 \pm 0.02$ for the BPE and HS methods respectively. Same format as Figure~\ref{J1048-5832_waterfall} otherwise.}
\label{J1453-6413_waterfall}
\end{figure*}

\begin{figure*}
\centering
\includegraphics[width = 1.0\textwidth]{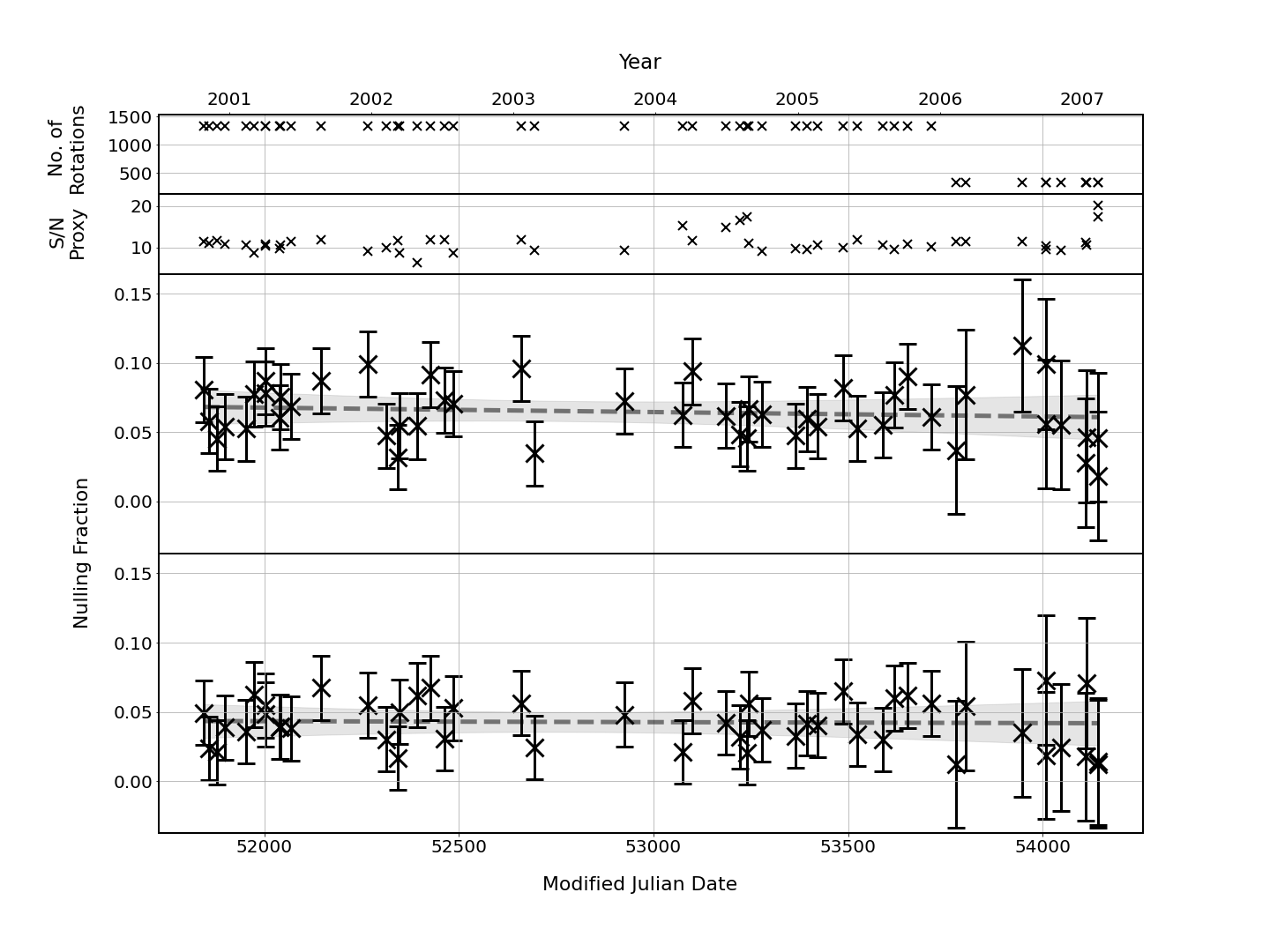}

\caption{Evolution of NF for PSR~J1453$-$6413. Same format as Figure~\ref{J1048-5832_nf_evolution} otherwise. }
\label{J1453-6413_nf_evolution}
\end{figure*}


\begin{figure*}
\centering
\includegraphics[width = 1.0\textwidth]{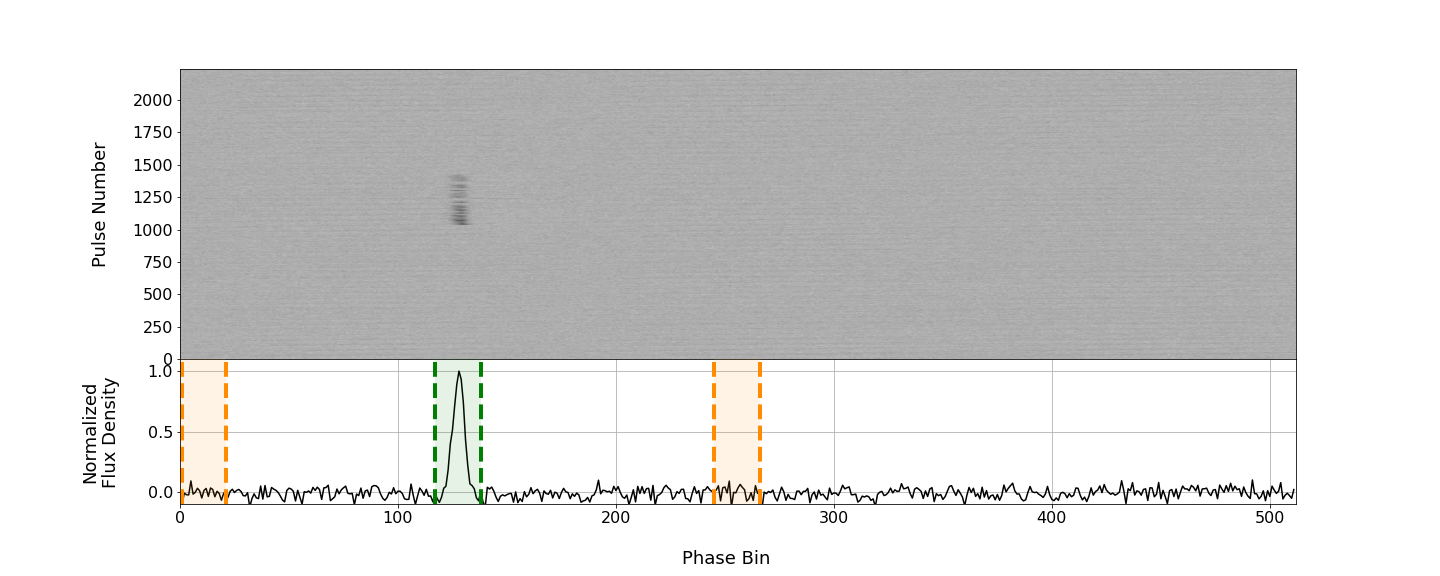}

\caption{Observation of PSR~J1502$-$5653 made on MJD 53147. The pulse profile has a single sharp and narrow Gaussian-like component, covering only a small fraction of the pulse period. The calculated NF for this observation was $0.91 \pm 0.02$ and $0.88 \pm 0.02$ for the BPE and HS methods respectively. Same format as Figure~\ref{J1048-5832_waterfall} otherwise.}
\label{J1502-5653_waterfall}
\end{figure*}

\begin{figure*}
\centering
\includegraphics[width = 1.0\textwidth]{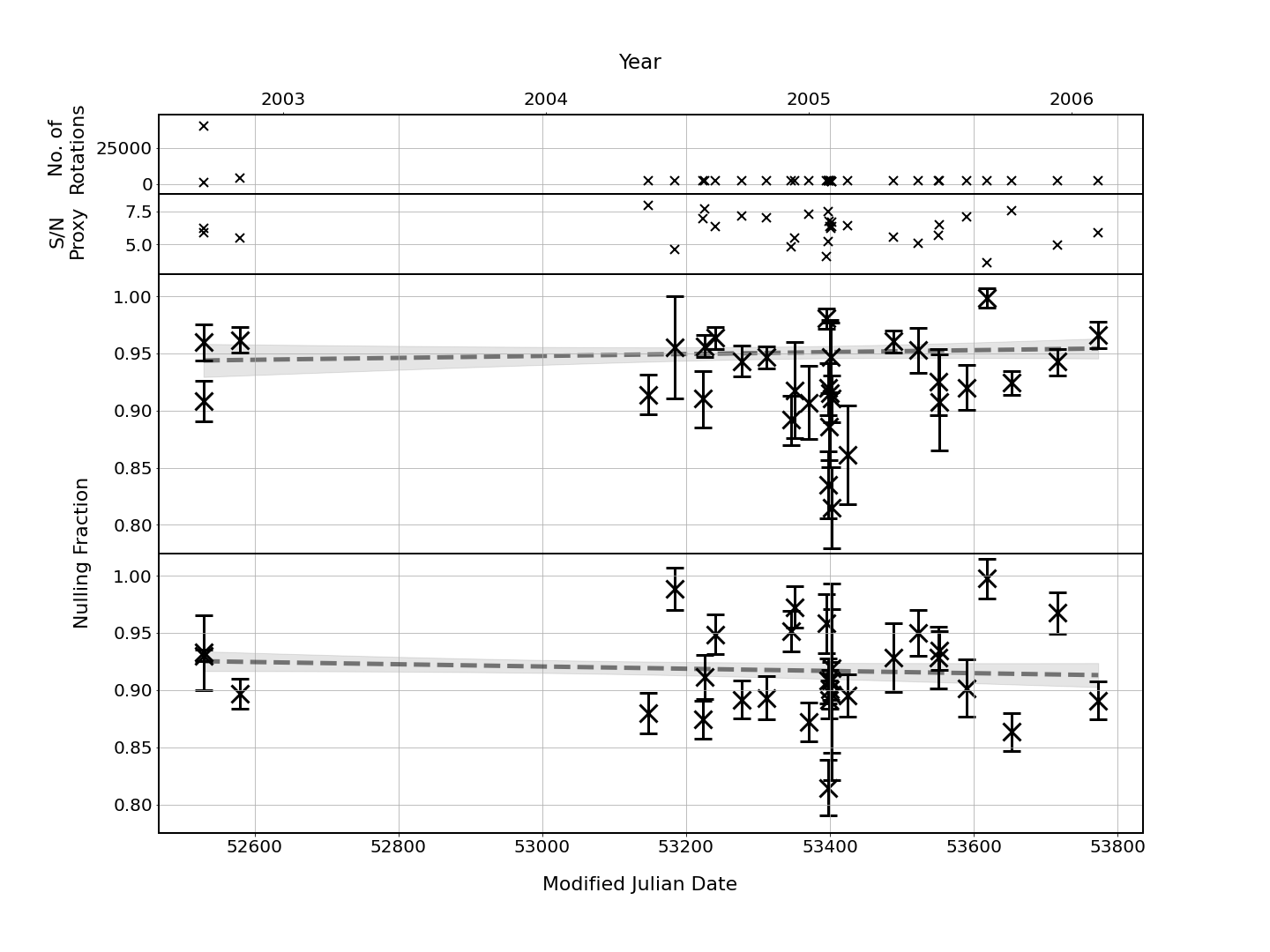}

\caption{Evolution of NF for PSR~J1502$-$5653. Same format as Figure~\ref{J1048-5832_nf_evolution} otherwise.}
\label{J1502-5653_nf_evolution}
\end{figure*}


\begin{figure*}
\centering
\includegraphics[width = 1.0\textwidth]{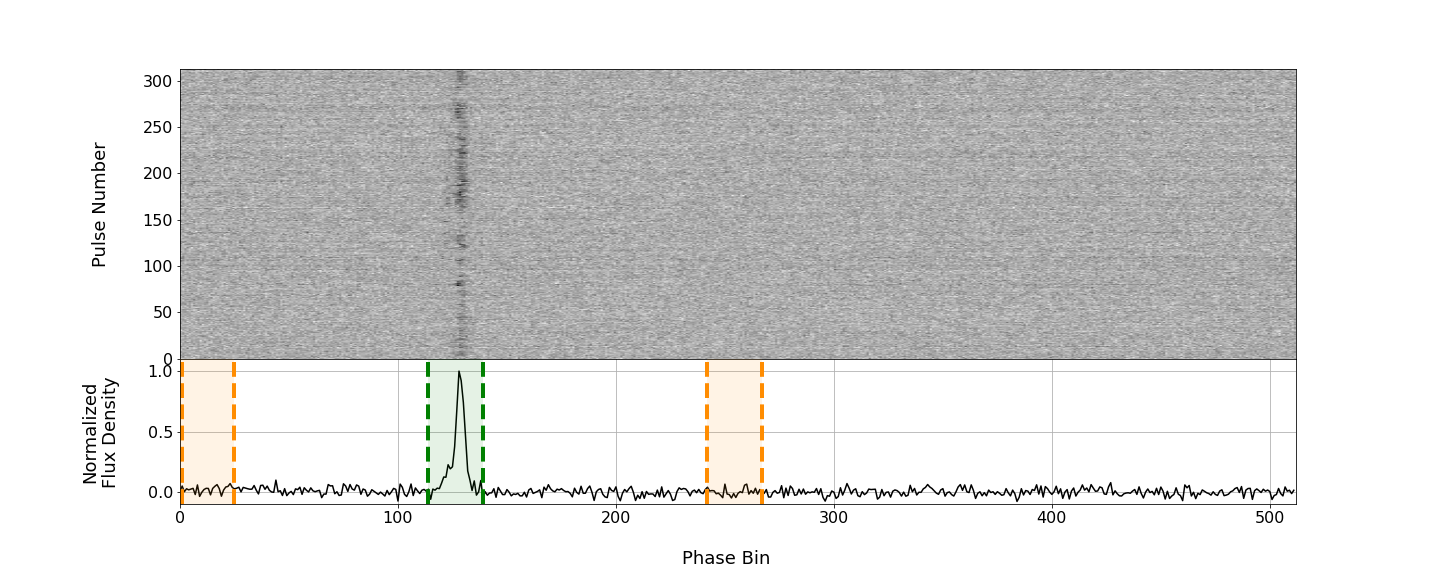}

\caption{Observation of PSR~J1559$-$5545 made on MJD 52069. The pulse profile appears to be comprised of two overlapping emission components. As a result the pulse profile is a sharp and narrow Gaussian-like component with a leading shoulder. The two components seem to null quasi-independently. The calculated NF for this observation was $0.30^{+0.07}_{-0.08}$
 and $0.38 \pm 0.04$ for the BPE and HS methods respectively. Same format as Figure~\ref{J1048-5832_waterfall} otherwise.}
\label{J1559-5545_waterfall}
\end{figure*}

\begin{figure*}
\centering
\includegraphics[width = 1.0\textwidth]{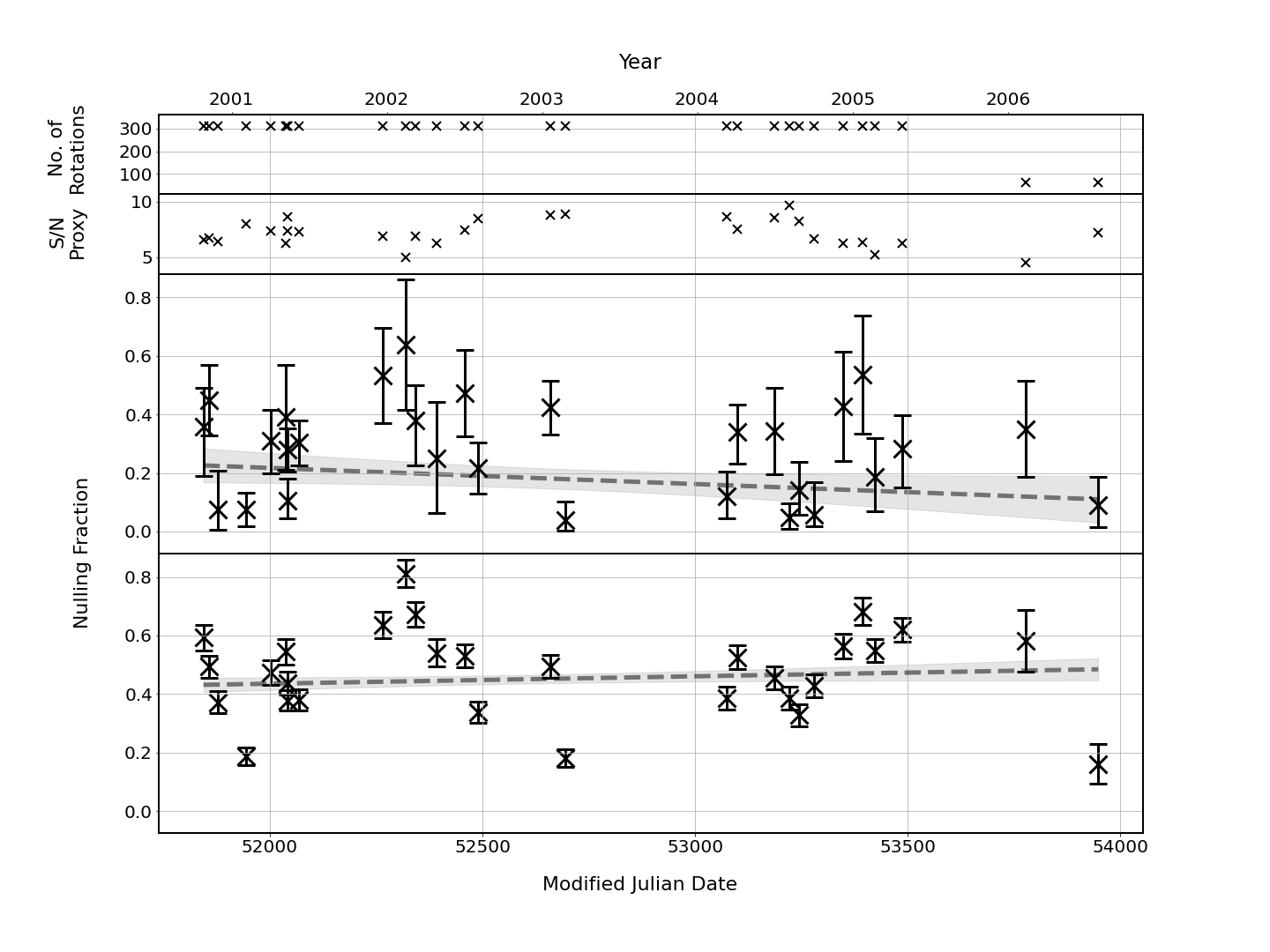}

\caption{Evolution of NF for PSR~J1559$-$5545. Same format as Figure~\ref{J1048-5832_nf_evolution} otherwise.}
\label{J1559-5545_nf_evolution}
\end{figure*}


\begin{figure*}
\centering
\includegraphics[width=\textwidth]{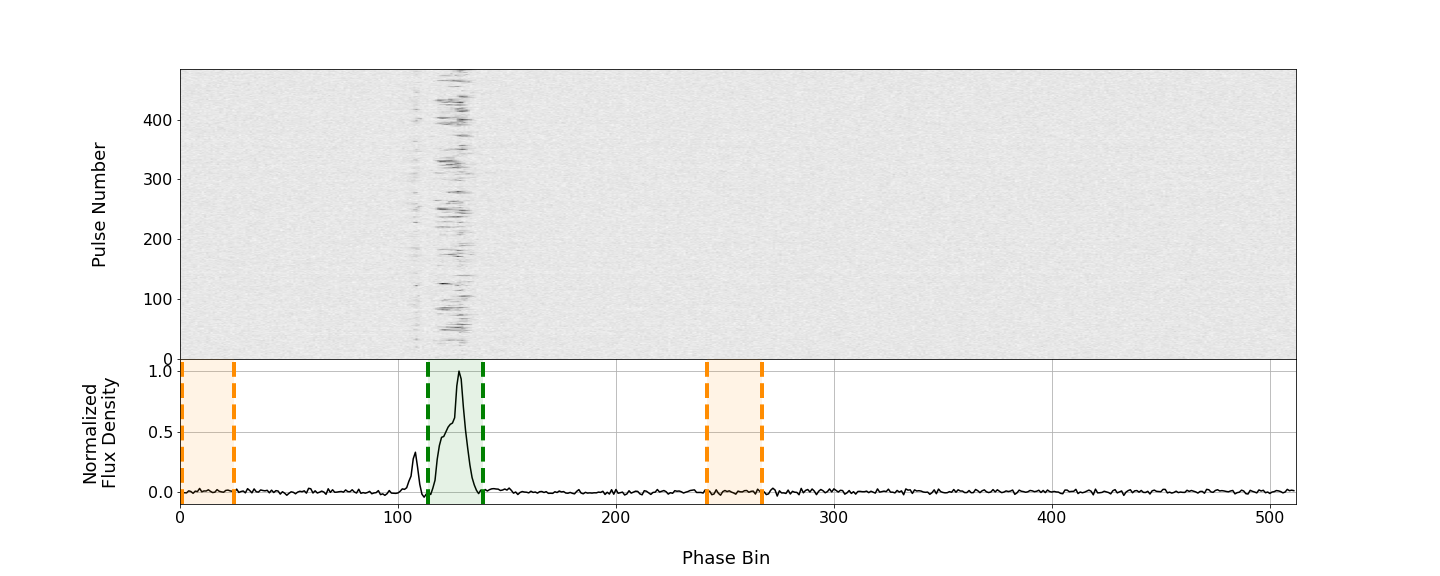}
\caption{Observation of PSR~J1745$-$3040 made on MJD 53367. The pulse profile consists of a small leading component and a more complex trailing feature composed of quasi-independent nulling components; here we focus on the latter. The calculated NF for this observation was $0.48 \pm 0.03$ and $0.47 \pm 0.04$ for the BPE and HS methods respectively. Same format as Figure~\ref{J1048-5832_waterfall} otherwise.}
\label{J1745-3040_major_waterfall}
\end{figure*}

\begin{figure*}
\centering
\includegraphics[width = 1.0\textwidth]{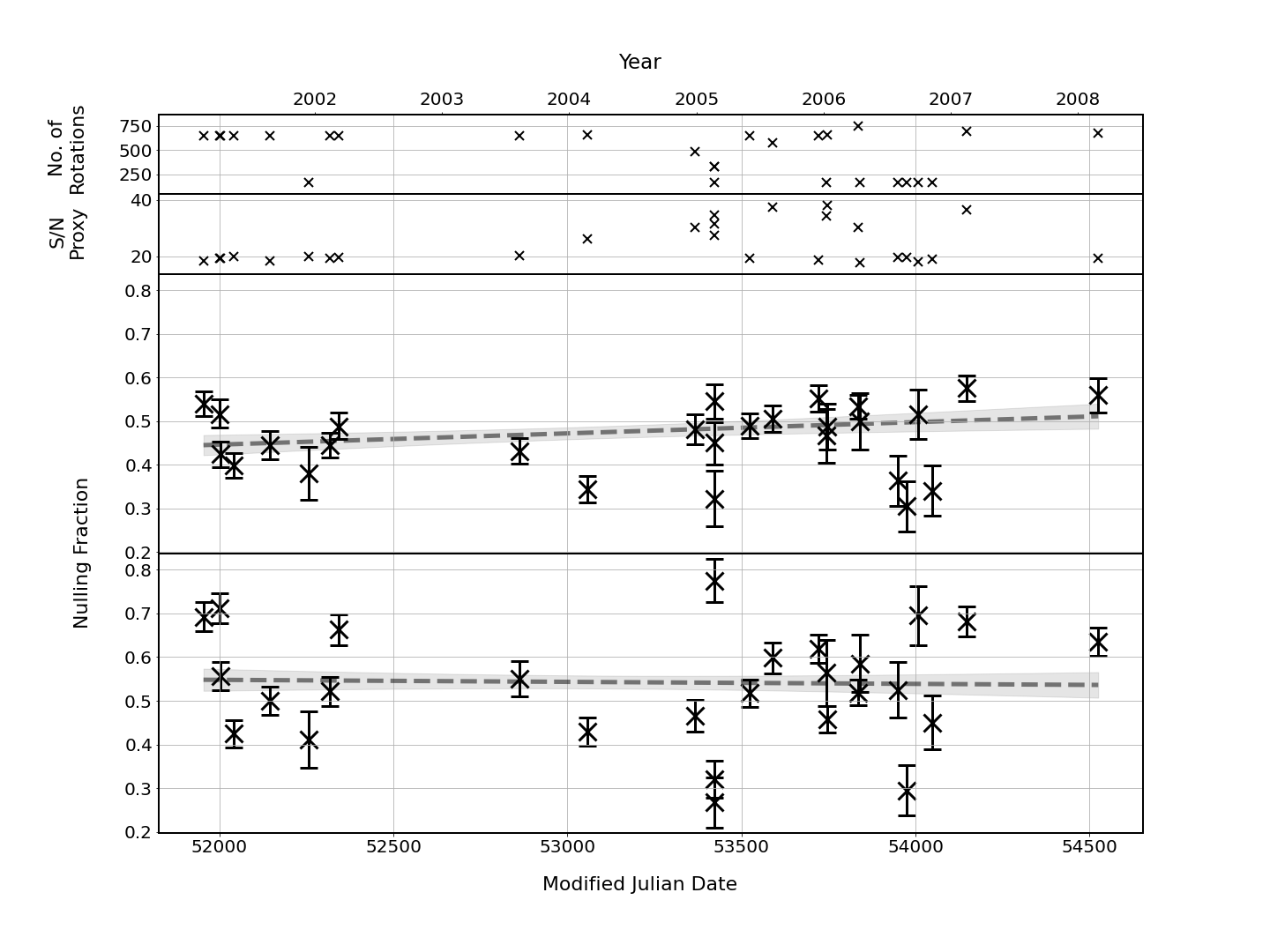}
\caption{Evolution of NF for main component of PSR~J1745$-$3040. Same format as Figure~\ref{J1048-5832_nf_evolution} otherwise.}
\label{J1745-3040_major_nf_evolution}
\end{figure*}

\begin{figure*}
\centering
\includegraphics[width=\textwidth]{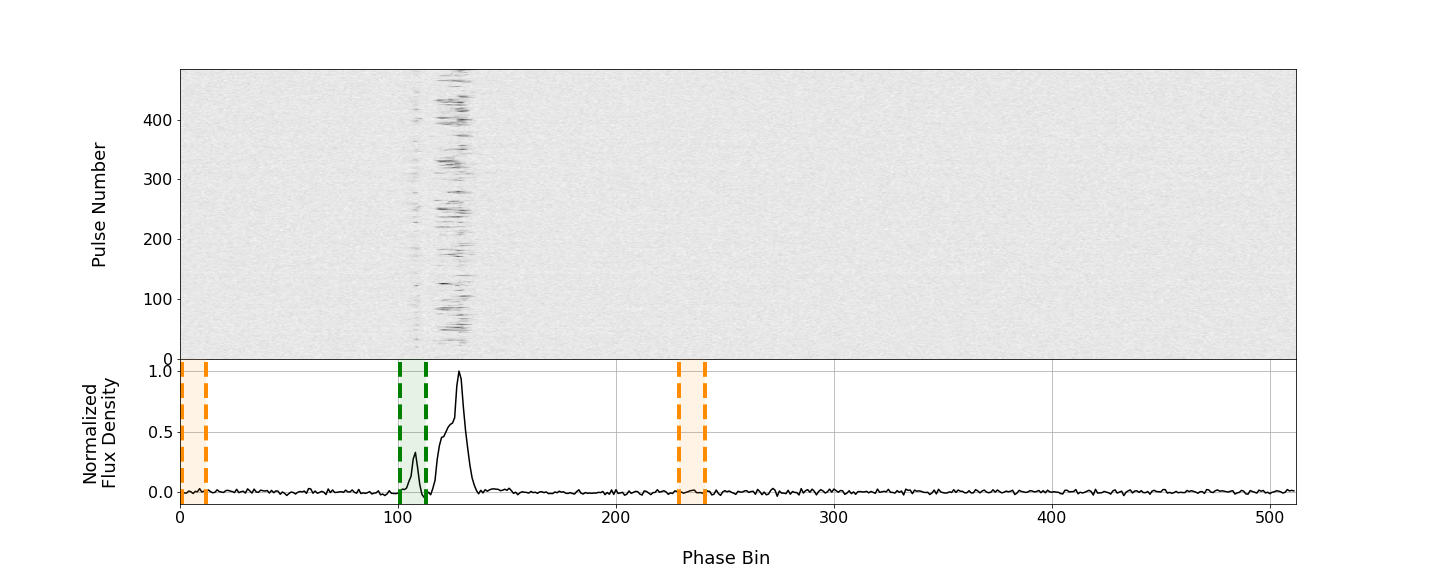}
\caption{Observation of PSR~J1745$-$3040 made on MJD 53367. The pulse profile consists of a small leading component and a more complex trailing feature composed of quasi-independent nulling components; here we focus on the former. The calculated NF for this observation was $0.53 \pm 0.04$ and $0.60 \pm 0.05$ for the BPE and HS methods respectively. Same format as Figure~\ref{J1048-5832_waterfall} otherwise.}
\label{J1745-3040_minor_waterfall}
\end{figure*}

\begin{figure*}
\centering
\includegraphics[width = 1.0\textwidth]{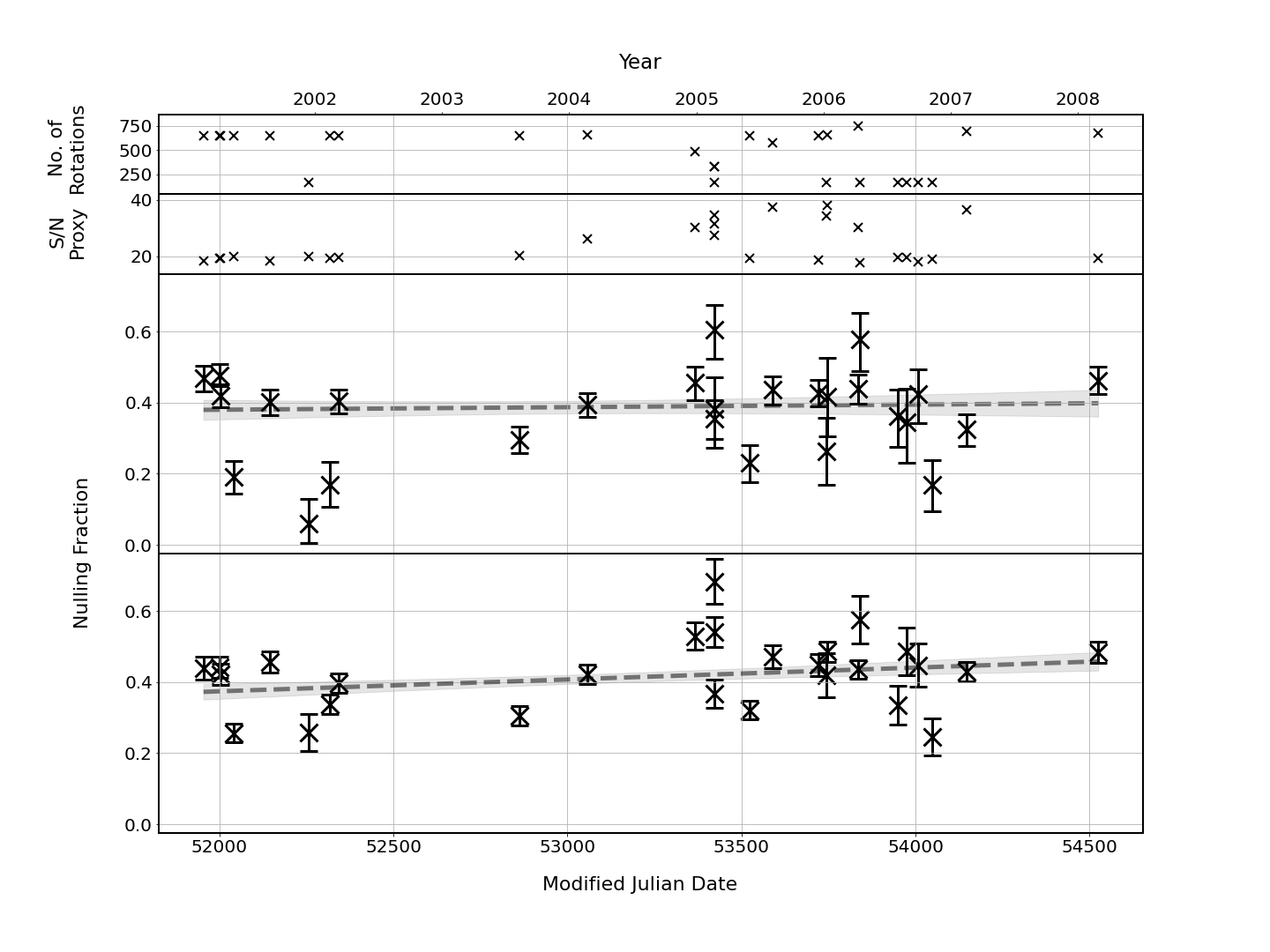}
\caption{Evolution of NF for minor component of PSR~J1745$-$3040. Same format as Figure~\ref{J1048-5832_nf_evolution} otherwise.}
\label{J1745-3040_minor_nf_evolution}
\end{figure*}


\begin{figure*}
\centering
\includegraphics[width = 1.0\textwidth]{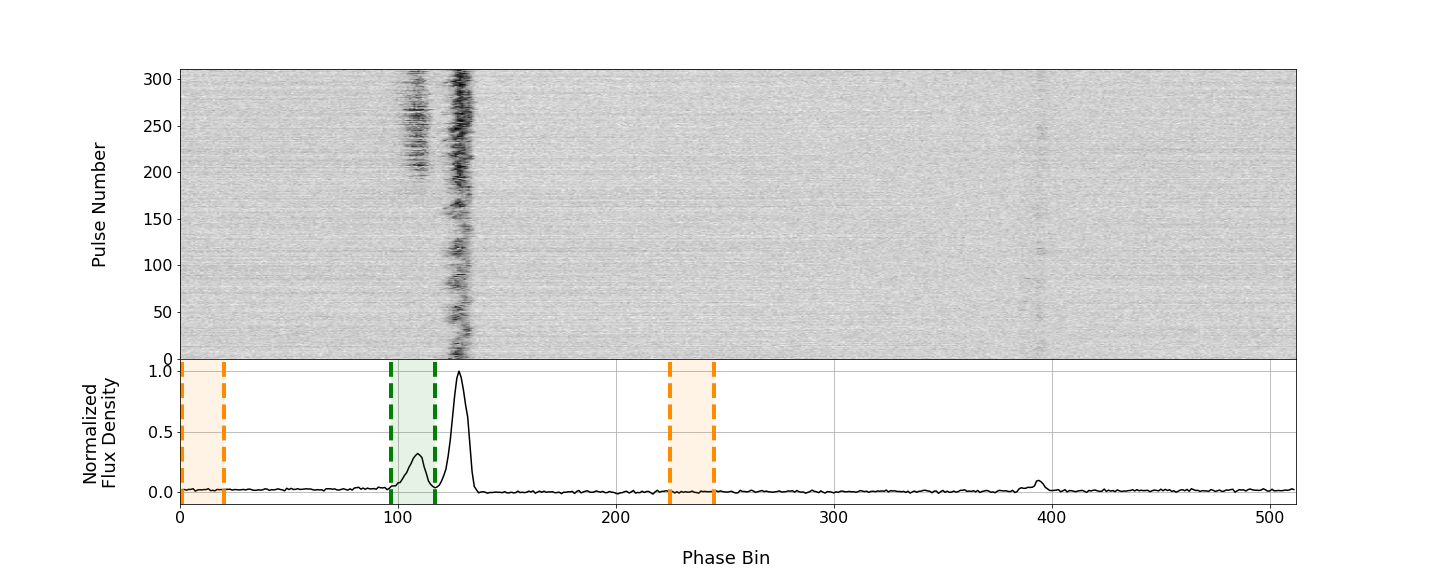}

\caption{Observation of PSR~J1825$-$0935 made on MJD 51844. There are three main components in the pulse profile: a main pulse, a precursor component leading the main pulse and an interpulse. The calculated NF for this observation was $0.58 \pm 0.05$ and $0.60 \pm 0.05$ for the BPE and HS methods respectively. Same format as Figure~\ref{J1048-5832_waterfall} otherwise.}
\label{J1825-0935_waterfall}
\end{figure*}

\begin{figure*}
\centering
\includegraphics[width = 1.0\textwidth]{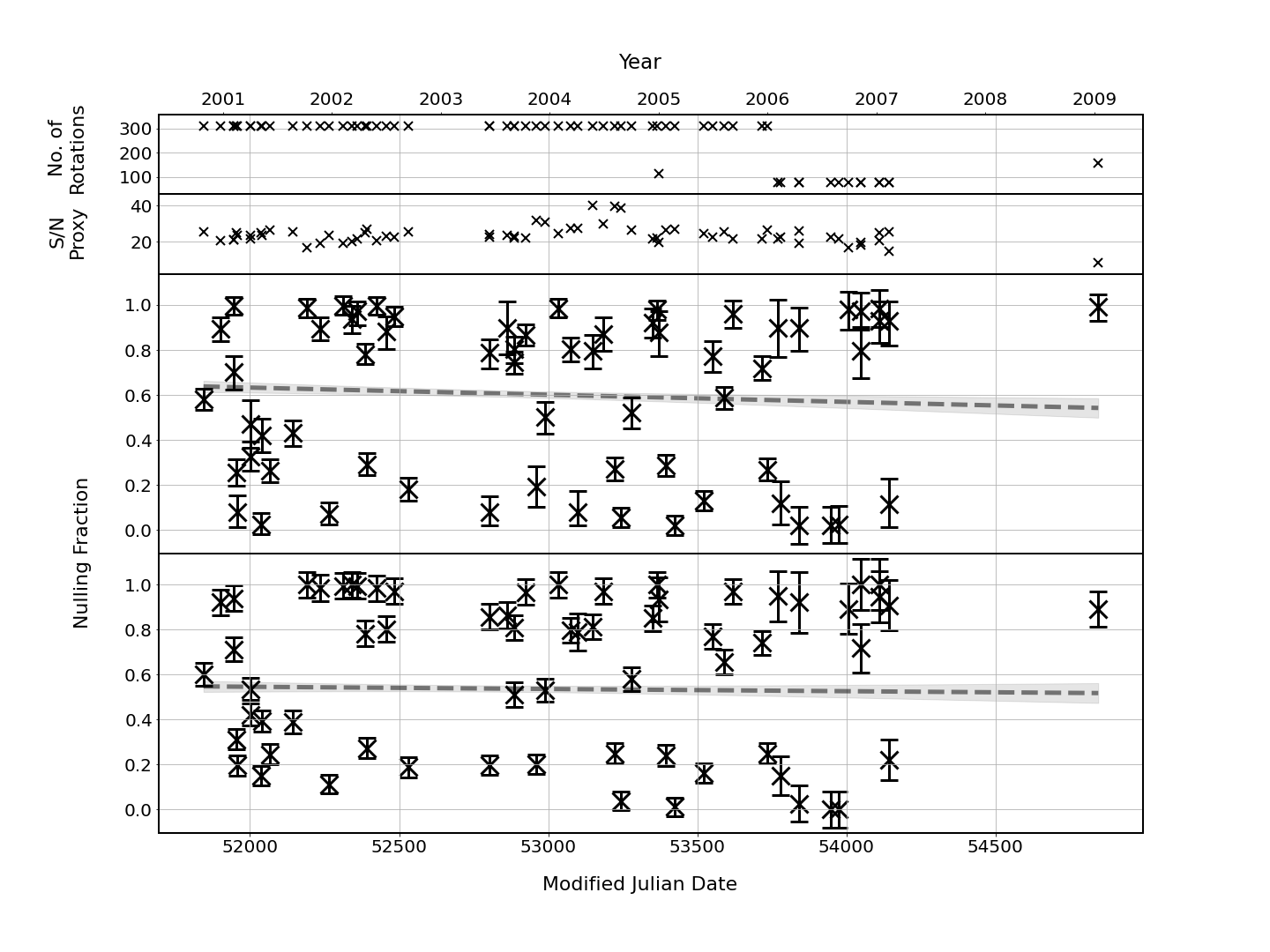}

\caption{Evolution of NF for PSR~J1825$-$0935. Same format as Figure~\ref{J1048-5832_nf_evolution} otherwise.}
\label{J1825-0935_nf_evolution}
\end{figure*}


\begin{figure*}
\centering
\includegraphics[width = 1.0\textwidth]{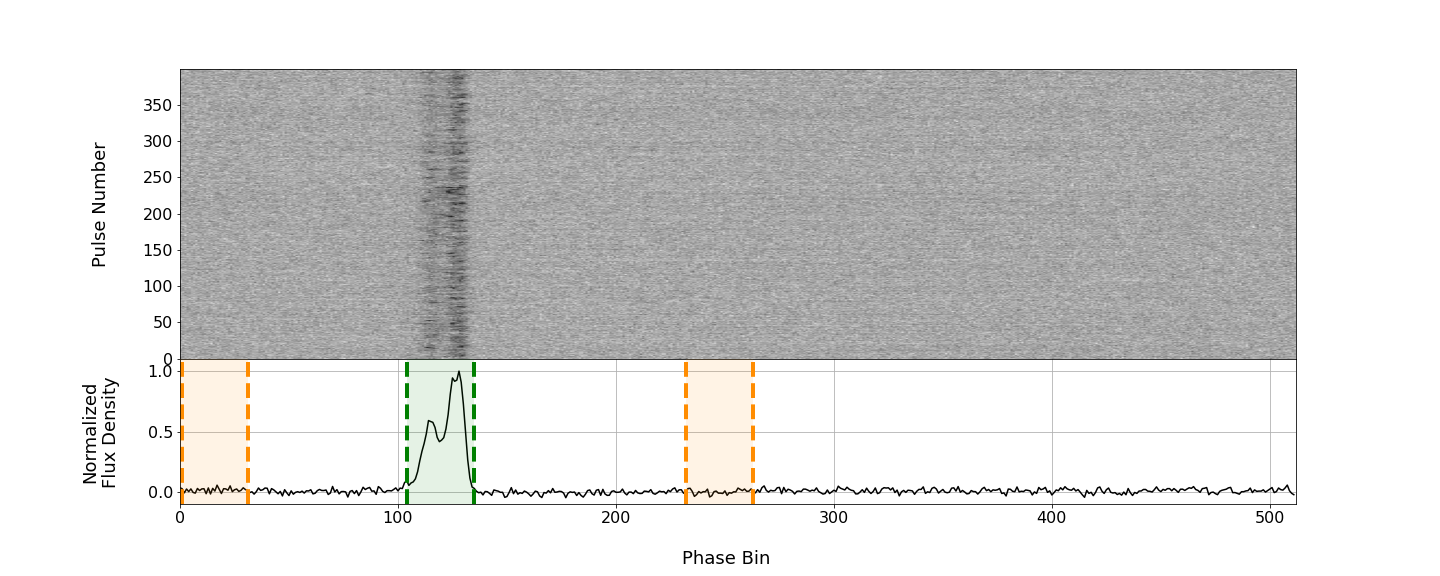}

\caption{Observation of PSR~J1847$-$0402 made on MJD 52069. The pulse profile of PSR~J1847$-$0402 consists of two overlapping Gaussian-like components, which appear to null quasi-independently. The calculated NF for this observation was $0.01 \pm 0.02$ and $0.02 \pm 0.02$ for the BPE and HS methods respectively. Same format as Figure~\ref{J1048-5832_waterfall} otherwise.}
\label{J1847-0402_waterfall}
\end{figure*}

\begin{figure*}
\centering
\includegraphics[width = 1.0\textwidth]{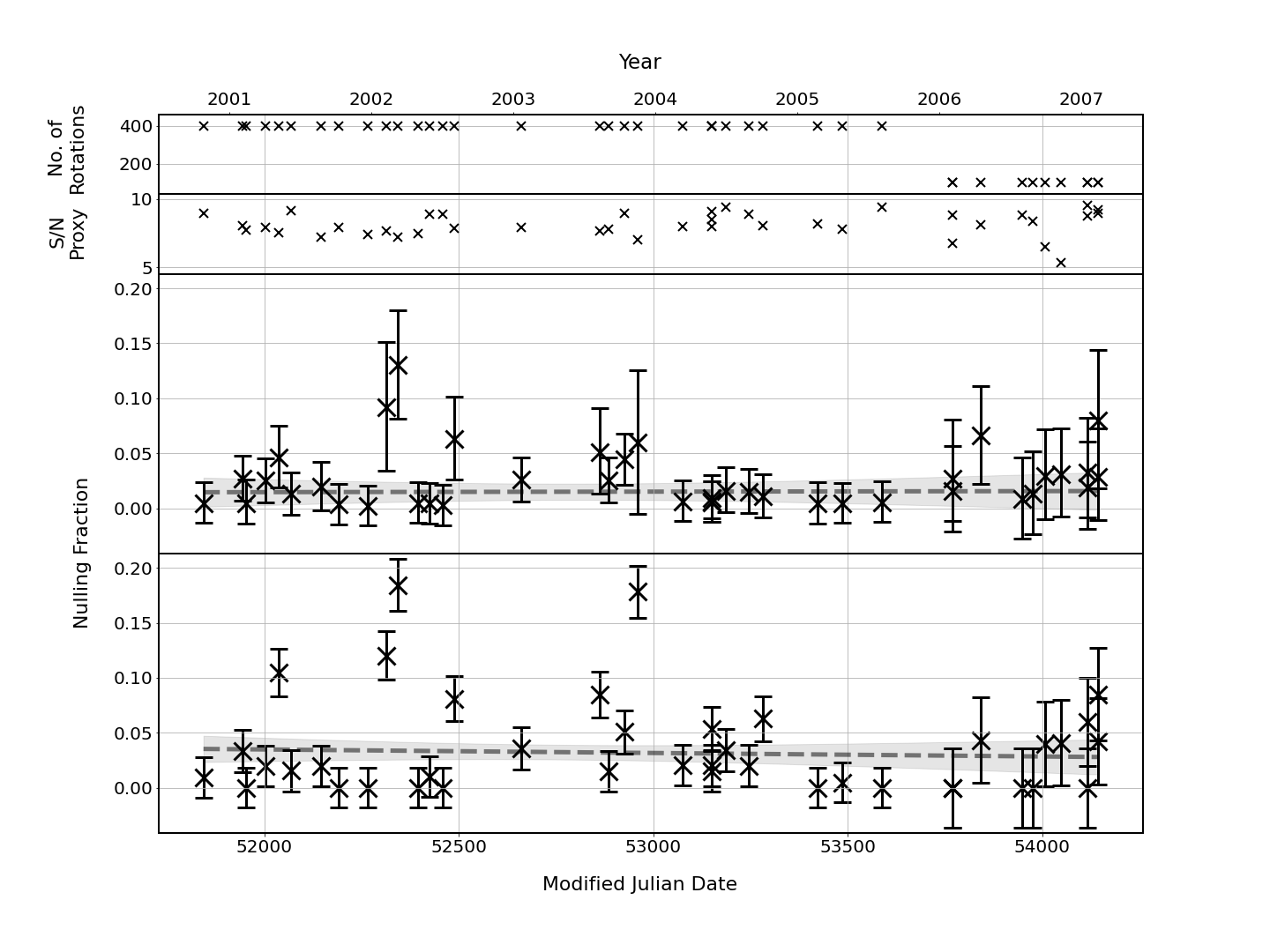}

\caption{Evolution of NF for PSR~J1847$-$0402. Same format as Figure~\ref{J1048-5832_nf_evolution} otherwise.}
\label{J1847-0402_nf_evolution}
\end{figure*}

\section*{Acknowledgments}

\emph{P.R.B. is supported by the UK Science and Technology Facilities Council (STFC), grant number ST/W000946/1. M.A.M. is a member of the NANOGrav Physics Frontiers Center, which is supported by the U.S. National Science Foundation (NSF) under award number 2020265. Murriyang, the Parkes 64 m radio telescope, is part of the Australia Telescope National Facility (\href{https://ror.org/05qajvd42}{https://ror.org/05qajvd42}), which is funded by the Australian
Government for operation as a National Facility managed by CSIRO.
}


\section*{Data availability}
The observational data used in this work are publicly available through the Parkes Pulsar Data Archive \citep{2011PASA...28..202H}, accessed via the CSIRO Data Access Portal. No new raw data were generated in the course of this study.

\bibliography{refs}

@ARTICLE{2007MNRAS.377.1383W,
           author = "{{Wang}, N. and {Manchester}, R.~N. and {Johnston}, S.}",
            title = "{Pulsar nulling and mode changing}",
          journal = {\mnras},
             year = "2007",
            month = "May",
           volume = {377},
           number = {3},
            pages = {1383-1392},
              doi = {10.1111/j.1365-2966.2007.11703.x},
    archivePrefix = {arXiv},
           eprint = {astro-ph/0703241},
     primaryClass = {astro-ph},
           adsurl = {https://ui.adsabs.harvard.edu/abs/2007MNRAS.377.1383W}
    }

@article{burke2012high,
      title={The High Time Resolution Universe Pulsar Survey--V. Single-pulse energetics and modulation properties of 315 pulsars},
      author={Burke-Spolaor, S and Johnston, S and Bailes, M and Bates, SD and Bhat, NDR and Burgay, M and Champion, DJ and D’Amico, Nicolo' and Keith, MJ and Kramer, M and others},
      journal={Monthly Notices of the Royal Astronomical Society},
      volume={423},
      number={2},
      pages={1351--1367},
      year={2012},
      publisher={Blackwell Publishing Ltd Oxford, UK}
    }

@ARTICLE{1976MNRAS.176..249R,
       author = {{Ritchings}, R.~T.},
        title = "{Pulsar single pulse intensity measurements and pulse nulling.}",
      journal = {\mnras},
         year = 1976,
        month = aug,
       volume = {176},
        pages = {249-263},
          doi = {10.1093/mnras/176.2.249},
       adsurl = {https://ui.adsabs.harvard.edu/abs/1976MNRAS.176..249R}
}

@article{10.1046/j.1365-8711.2000.03713.x,
    author = {Wang, N. and Manchester, R. N. and Pace, R. T. and Bailes, M. and Kaspi, V. M. and Stappers, B. W. and Lyne, A. G.},
    title = "{Glitches in southern pulsars}",
    journal = {Monthly Notices of the Royal Astronomical Society},
    volume = {317},
    number = {4},
    pages = {843-860},
    year = {2000},
    month = {10},
    issn = {0035-8711},
    doi = {10.1046/j.1365-8711.2000.03713.x},
    url = {https://doi.org/10.1046/j.1365-8711.2000.03713.x},
    eprint = {https://academic.oup.com/mnras/article-pdf/317/4/843/3492365/317-4-843.pdf},
}

@article{10.1093/mnras/255.3.401,
    author = {Johnston, Simon and Lyne, A. G. and Manchester, R. N. and Kniffen, D. A. and D'Amico, N. and Lim, J. and Ashworth, M.},
    title = "{A high-frequency survey of the southern Galactic plane for pulsars}",
    journal = {Monthly Notices of the Royal Astronomical Society},
    volume = {255},
    number = {3},
    pages = {401-411},
    year = {1992},
    month = {04},
    issn = {0035-8711},
    doi = {10.1093/mnras/255.3.401},
    url = {https://doi.org/10.1093/mnras/255.3.401},
    eprint = {https://academic.oup.com/mnras/article-pdf/255/3/401/3366209/mnras255-0401.pdf},
}

@ARTICLE{2010ApJ...713..154A,
       author = {{Abdo}, A.~A. and {Ackermann}, M. and {Ajello}, M. and {Allafort}, A. and {Atwood}, W.~B. and {Baldini}, L. and {Ballet}, J. and {Barbiellini}, G. and {Baring}, M.~G. and {Bastieri}, D. and {Baughman}, B.~M. and {Bechtol}, K. and {Bellazzini}, R. and {Berenji}, B. and {Blandford}, R.~D. and {Bloom}, E.~D. and {Bonamente}, E. and {Borgland}, A.~W. and {Bouvier}, A. and {Bregeon}, J. and {Brez}, A. and {Brigida}, M. and {Bruel}, P. and {Burnett}, T.~H. and {Buson}, S. and {Caliandro}, G.~A. and {Cameron}, R.~A. and {Caraveo}, P.~A. and {Carrigan}, S. and {Casandjian}, J.~M. and {Cecchi}, C. and {{\c{C}}elik}, {\"O}. and {Chekhtman}, A. and {Cheung}, C.~C. and {Chiang}, J. and {Ciprini}, S. and {Claus}, R. and {Cohen-Tanugi}, J. and {Conrad}, J. and {Dermer}, C.~D. and {de Luca}, A. and {de Palma}, F. and {Dormody}, M. and {Silva}, E. do Couto e. and {Drell}, P.~S. and {Dubois}, R. and {Dumora}, D. and {Farnier}, C. and {Favuzzi}, C. and {Fegan}, S.~J. and {Focke}, W.~B. and {Fortin}, P. and {Frailis}, M. and {Fukazawa}, Y. and {Funk}, S. and {Fusco}, P. and {Gargano}, F. and {Gasparrini}, D. and {Gehrels}, N. and {Germani}, S. and {Giavitto}, G. and {Giebels}, B. and {Giglietto}, N. and {Giordano}, F. and {Glanzman}, T. and {Godfrey}, G. and {Grenier}, I.~A. and {Grondin}, M. -H. and {Grove}, J.~E. and {Guillemot}, L. and {Guiriec}, S. and {Hadasch}, D. and {Harding}, A.~K. and {Hays}, E. and {Hobbs}, G. and {Horan}, D. and {Hughes}, R.~E. and {Jackson}, M.~S. and {J{\'o}hannesson}, G. and {Johnson}, A.~S. and {Johnson}, T.~J. and {Johnson}, W.~N. and {Kamae}, T. and {Katagiri}, H. and {Kataoka}, J. and {Kawai}, N. and {Kerr}, M. and {Kn{\"o}dlseder}, J. and {Kuss}, M. and {Lande}, J. and {Latronico}, L. and {Lee}, S. -H. and {Lemoine-Goumard}, M. and {Llena Garde}, M. and {Longo}, F. and {Loparco}, F. and {Lott}, B. and {Lovellette}, M.~N. and {Lubrano}, P. and {Makeev}, A. and {Manchester}, R.~N. and {Marelli}, M. and {Mazziotta}, M.~N. and {McConville}, W. and {McEnery}, J.~E. and {McGlynn}, S. and {Meurer}, C. and {Michelson}, P.~F. and {Mitthumsiri}, W. and {Mizuno}, T. and {Moiseev}, A.~A. and {Monte}, C. and {Monzani}, M.~E. and {Morselli}, A. and {Moskalenko}, I.~V. and {Murgia}, S. and {Nakamori}, T. and {Nolan}, P.~L. and {Norris}, J.~P. and {Noutsos}, A. and {Nuss}, E. and {Ohsugi}, T. and {Omodei}, N. and {Orlando}, E. and {Ormes}, J.~F. and {Ozaki}, M. and {Paneque}, D. and {Panetta}, J.~H. and {Parent}, D. and {Pelassa}, V. and {Pepe}, M. and {Pesce-Rollins}, M. and {Pierbattista}, M. and {Piron}, F. and {Porter}, T.~A. and {Rain{\`o}}, S. and {Rando}, R. and {Ray}, P.~S. and {Razzano}, M. and {Reimer}, A. and {Reimer}, O. and {Reposeur}, T. and {Ritz}, S. and {Rochester}, L.~S. and {Rodriguez}, A.~Y. and {Romani}, R.~W. and {Roth}, M. and {Ryde}, F. and {Sadrozinski}, H.~F. -W. and {Sander}, A. and {Saz Parkinson}, P.~M. and {Sgr{\`o}}, C. and {Siskind}, E.~J. and {Smith}, D.~A. and {Smith}, P.~D. and {Spandre}, G. and {Spinelli}, P. and {Starck}, J. -L. and {Strickman}, M.~S. and {Suson}, D.~J. and {Takahashi}, H. and {Takahashi}, T. and {Tanaka}, T. and {Thayer}, J.~B. and {Thayer}, J.~G. and {Thompson}, D.~J. and {Tibaldo}, L. and {Torres}, D.~F. and {Tosti}, G. and {Tramacere}, A. and {Usher}, T.~L. and {Van Etten}, A. and {Vasileiou}, V. and {Venter}, C. and {Vilchez}, N. and {Vitale}, V. and {Waite}, A.~P. and {Wang}, P. and {Watters}, K. and {Weltevrede}, P. and {Winer}, B.~L. and {Wood}, K.~S. and {Ylinen}, T. and {Ziegler}, M.},
        title = "{The Vela Pulsar: Results from the First Year of Fermi LAT Observations}",
      journal = {\apj},
         year = 2010,
        month = apr,
       volume = {713},
       number = {1},
        pages = {154-165},
          doi = {10.1088/0004-637X/713/1/154},
archivePrefix = {arXiv},
       eprint = {1002.4050},
 primaryClass = {astro-ph.HE},
       adsurl = {https://ui.adsabs.harvard.edu/abs/2010ApJ...713..154A}
}

@article{Gonzalez_2006,
doi = {10.1086/507125},
url = {https://dx.doi.org/10.1086/507125},
year = {2006},
month = {nov},
publisher = {},
volume = {652},
number = {1},
pages = {569},
author = {M. E. Gonzalez and V. M. Kaspi and M. J. Pivovaroff and B. M. Gaensler},
title = {Chandra and XMM-Newton Observations of the Vela-like Pulsar B1046–58},
journal = {The Astrophysical Journal}
}

@article{10.1093/mnras/stab095,
    author = {Johnston, Simon and Sobey, C and Dai, S and Keith, M and Kerr, M and Manchester, R N and Oswald, L S and Parthasarathy, A and Shannon, R M and Weltevrede, P},
    title = "{Two years of pulsar observations with the ultra-wide-band receiver on the Parkes radio telescope}",
    journal = {Monthly Notices of the Royal Astronomical Society},
    volume = {502},
    number = {1},
    pages = {1253-1262},
    year = {2021},
    month = {01},
    issn = {0035-8711},
    doi = {10.1093/mnras/stab095},
    url = {https://doi.org/10.1093/mnras/stab095},
    eprint = {https://academic.oup.com/mnras/article-pdf/502/1/1253/36179828/stab095.pdf},
}

@ARTICLE{2019MNRAS.489.3810P,
       author = {{Parthasarathy}, A. and {Shannon}, R.~M. and {Johnston}, S. and {Lentati}, L. and {Bailes}, M. and {Dai}, S. and {Kerr}, M. and {Manchester}, R.~N. and {Os{\l}owski}, S. and {Sobey}, C. and {van Straten}, W. and {Weltevrede}, P.},
        title = "{Timing of young radio pulsars - I. Timing noise, periodic modulation, and proper motion}",
      journal = {\mnras},
         year = 2019,
        month = nov,
       volume = {489},
       number = {3},
        pages = {3810-3826},
          doi = {10.1093/mnras/stz2383},
archivePrefix = {arXiv},
       eprint = {1908.11709},
 primaryClass = {astro-ph.HE},
       adsurl = {https://ui.adsabs.harvard.edu/abs/2019MNRAS.489.3810P}
}

@ARTICLE{2023RAA....23j5014L,
       author = {{Li}, Wei and {Dang}, Shi-Jun and {Yuan}, Jian-Ping and {Li}, Lin and {Wang}, Wei-Hua and {Shang}, Lun-Hua and {Wang}, Na and {Li}, Qing-Ying and {Lu}, Ji-Guang and {Kou}, Fei-Fei and {Wang}, Shuang-Qiang and {Xiao}, Shuo and {Zhi}, Qi-Jun and {Liu}, Yu-Lan and {Zhao}, Ru-Shuang and {Dong}, Ai-Jun and {Zhang}, Bin and {You}, Zi-Yi and {Cai}, Yan-Qing and {Yang}, Ya-Qin and {Ren}, Ying-Ying and {Liu}, Yu-Jia and {Xu}, Heng},
        title = "{Results of 23 yr of Pulsar Timing of PSR J1453-6413}",
      journal = {Research in Astronomy and Astrophysics},
         year = 2023,
        month = oct,
       volume = {23},
       number = {10},
          eid = {105014},
        pages = {105014},
          doi = {10.1088/1674-4527/acf1e1},
archivePrefix = {arXiv},
       eprint = {2306.13611},
 primaryClass = {astro-ph.HE},
       adsurl = {https://ui.adsabs.harvard.edu/abs/2023RAA....23j5014L}
}

@ARTICLE{2021MNRAS.502.1253J,
       author = {{Johnston}, Simon and {Sobey}, C. and {Dai}, S. and {Keith}, M. and {Kerr}, M. and {Manchester}, R.~N. and {Oswald}, L.~S. and {Parthasarathy}, A. and {Shannon}, R.~M. and {Weltevrede}, P.},
        title = "{Two years of pulsar observations with the ultra-wide-band receiver on the Parkes radio telescope}",
      journal = {\mnras},
         year = 2021,
        month = mar,
       volume = {502},
       number = {1},
        pages = {1253-1262},
          doi = {10.1093/mnras/stab095},
archivePrefix = {arXiv},
       eprint = {2101.07373},
 primaryClass = {astro-ph.HE},
       adsurl = {https://ui.adsabs.harvard.edu/abs/2021MNRAS.502.1253J}
}

@ARTICLE{2010MNRAS.404...30B,
       author = {{Backus}, Isaac and {Mitra}, Dipanjan and {Rankin}, Joanna M.},
        title = "{Dynamic emission properties of pulsars B0943+10 and B1822-09 - I. Comparison, and the discovery of a `Q'-mode precursor}",
      journal = {\mnras},
         year = 2010,
        month = may,
       volume = {404},
       number = {1},
        pages = {30-41},
          doi = {10.1111/j.1365-2966.2009.16102.x},
       adsurl = {https://ui.adsabs.harvard.edu/abs/2010MNRAS.404...30B}
}

@ARTICLE{1982A&A...109..279F,
       author = {{Fowler}, L.~A. and {Wright}, G.~A.~E.},
        title = "{Pulse-interpulse interaction in pulsar PSR 1822-09.}",
      journal = {\aap},
         year = 1982,
        month = may,
       volume = {109},
        pages = {279-281},
       adsurl = {https://ui.adsabs.harvard.edu/abs/1982A&A...109..279F}
}

@ARTICLE{2012MNRAS.427..180L,
       author = {{Latham}, Crystal and {Mitra}, Dipanjan and {Rankin}, Joanna},
        title = "{Modal sequencing and dynamic emission properties of an 8-h Giant Metrewave Radio Telescope observation of pulsar B1822-09}",
      journal = {\mnras},
         year = 2012,
        month = nov,
       volume = {427},
       number = {1},
        pages = {180-189},
          doi = {10.1111/j.1365-2966.2012.21985.x},
archivePrefix = {arXiv},
       eprint = {1209.1623},
 primaryClass = {astro-ph.SR},
       adsurl = {https://ui.adsabs.harvard.edu/abs/2012MNRAS.427..180L}
}

@ARTICLE{2001MNRAS.328...17M,
       author = {{Manchester}, R.~N. and {Lyne}, A.~G. and {Camilo}, F. and {Bell}, J.~F. and {Kaspi}, V.~M. and {D'Amico}, N. and {McKay}, N.~P.~F. and {Crawford}, F. and {Stairs}, I.~H. and {Possenti}, A. and {Kramer}, M. and {Sheppard}, D.~C.},
        title = "{The Parkes multi-beam pulsar survey - I. Observing and data analysis systems, discovery and timing of 100 pulsars}",
      journal = {\mnras},
         year = 2001,
        month = nov,
       volume = {328},
       number = {1},
        pages = {17-35},
          doi = {10.1046/j.1365-8711.2001.04751.x},
archivePrefix = {arXiv},
       eprint = {astro-ph/0106522},
 primaryClass = {astro-ph},
       adsurl = {https://ui.adsabs.harvard.edu/abs/2001MNRAS.328...17M}
}

@ARTICLE{1970Natur.228.1297B,
       author = {{Backer}, D.~C.},
        title = "{Peculiar Pulse Burst in PSR 1237 + 25}",
      journal = {\nat},
         year = 1970,
        month = dec,
       volume = {228},
       number = {5278},
        pages = {1297-1298},
          doi = {10.1038/2281297a0},
       adsurl = {https://ui.adsabs.harvard.edu/abs/1970Natur.228.1297B}
}

@ARTICLE{1970Natur.228...42B,
       author = {{Backer}, D.~C.},
        title = "{Pulsar Nulling Phenomena}",
      journal = {\nat},
         year = 1970,
        month = oct,
       volume = {228},
       number = {5266},
        pages = {42-43},
          doi = {10.1038/228042a0},
       adsurl = {https://ui.adsabs.harvard.edu/abs/1970Natur.228...42B}
}

@ARTICLE{1992ApJ...394..574B,
       author = {{Biggs}, James D.},
        title = "{An Analysis of Radio Pulsar Nulling Statistics}",
      journal = {\apj},
         year = 1992,
        month = aug,
       volume = {394},
        pages = {574},
          doi = {10.1086/171608},
       adsurl = {https://ui.adsabs.harvard.edu/abs/1992ApJ...394..574B}
}

@ARTICLE{2006Natur.439..817M,
       author = {{McLaughlin}, M.~A. and {Lyne}, A.~G. and {Lorimer}, D.~R. and {Kramer}, M. and {Faulkner}, A.~J. and {Manchester}, R.~N. and {Cordes}, J.~M. and {Camilo}, F. and {Possenti}, A. and {Stairs}, I.~H. and {Hobbs}, G. and {D'Amico}, N. and {Burgay}, M. and {O'Brien}, J.~T.},
        title = "{Transient radio bursts from rotating neutron stars}",
      journal = {\nat},
         year = 2006,
        month = feb,
       volume = {439},
       number = {7078},
        pages = {817-820},
          doi = {10.1038/nature04440},
archivePrefix = {arXiv},
       eprint = {astro-ph/0511587},
 primaryClass = {astro-ph},
       adsurl = {https://ui.adsabs.harvard.edu/abs/2006Natur.439..817M}
}

@article{Manchester2005,
  author = {R. N. Manchester and others},
  title = {The ATNF Pulsar Catalogue},
  journal = {The Astrophysical Journal},
  volume = {129},
  number = {6},
  pages = {1993--2006},
  year = {2005},
  doi = {10.1086/429647},
}

@article{Yan_2019,
   title={Periodic mode changing in PSR J1048−5832},
   volume={491},
   ISSN={1365-2966},
   url={http://dx.doi.org/10.1093/mnras/stz3399},
   DOI={10.1093/mnras/stz3399},
   number={4},
   journal={Monthly Notices of the Royal Astronomical Society},
   publisher={Oxford University Press (OUP)},
   author={Yan, W M and Manchester, R N and Wang, N and Wen, Z G and Yuan, J P and Lee, K J and Chen, J L},
   year={2019},
   month=dec, pages={4634–4641} }

@ARTICLE{1970Natur.227.1123D,
       author = {{Davies}, J.~G. and {Large}, M.~I. and {Pickwick}, A.~C.},
        title = "{Five New Pulsars}",
      journal = {\nat},
         year = 1970,
        month = sep,
       volume = {227},
       number = {5263},
        pages = {1123-1124},
          doi = {10.1038/2271123a0},
       adsurl = {https://ui.adsabs.harvard.edu/abs/1970Natur.227.1123D}
}

@article{10.1093/mnras/stab3336,
    author = {Basu, A and Shaw, B and Antonopoulou, D and Keith, M J and Lyne, A G and Mickaliger, M B and Stappers, B W and Weltevrede, P and Jordan, C A},
    title = {The Jodrell bank glitch catalogue: 106 new rotational glitches in 70 pulsars},
    journal = {Monthly Notices of the Royal Astronomical Society},
    volume = {510},
    number = {3},
    pages = {4049-4062},
    year = {2021},
    month = {11},
    issn = {0035-8711},
    doi = {10.1093/mnras/stab3336},
    url = {https://doi.org/10.1093/mnras/stab3336},
    eprint = {https://academic.oup.com/mnras/article-pdf/510/3/4049/42194660/stab3336.pdf},
}

@article{10.1093/mnras/stab2678,
    author = {Lower, M E and Johnston, S and Dunn, L and Shannon, R M and Bailes, M and Dai, S and Kerr, M and Manchester, R N and Melatos, A and Oswald, L S and Parthasarathy, A and Sobey, C and Weltevrede, P},
    title = {The impact of glitches on young pulsar rotational evolution},
    journal = {Monthly Notices of the Royal Astronomical Society},
    volume = {508},
    number = {3},
    pages = {3251-3274},
    year = {2021},
    month = {09},
    issn = {0035-8711},
    doi = {10.1093/mnras/stab2678},
    url = {https://doi.org/10.1093/mnras/stab2678},
    eprint = {https://academic.oup.com/mnras/article-pdf/508/3/3251/40736018/stab2678.pdf},
}

@ARTICLE{2018ApJ...855...14K,
       author = {{Kaplan}, D.~L. and {Swiggum}, J.~K. and {Fichtenbauer}, T.~D.~J. and {Vallisneri}, M.},
        title = "{A Gaussian Mixture Model for Nulling Pulsars}",
      journal = {\apj},
         year = 2018,
        month = mar,
       volume = {855},
       number = {1},
          eid = {14},
        pages = {14},
          doi = {10.3847/1538-4357/aaab62},
archivePrefix = {arXiv},
       eprint = {1801.09598},
 primaryClass = {astro-ph.IM},
       adsurl = {https://ui.adsabs.harvard.edu/abs/2018ApJ...855...14K}
}

@article{10.1093/mnras/stab282,
    author = {Sheikh, Sofia Z and MacDonald, Mariah G},
    title = {A statistical analysis of the nulling pulsar population},
    journal = {Monthly Notices of the Royal Astronomical Society},
    volume = {502},
    number = {4},
    pages = {4669-4679},
    year = {2021},
    month = {02},
    issn = {0035-8711},
    doi = {10.1093/mnras/stab282},
    url = {https://doi.org/10.1093/mnras/stab282},
    eprint = {https://academic.oup.com/mnras/article-pdf/502/4/4669/36392370/stab282.pdf},
}

@ARTICLE{2006Sci...312..549K,
       author = {{Kramer}, M. and {Lyne}, A.~G. and {O'Brien}, J.~T. and {Jordan}, C.~A. and {Lorimer}, D.~R.},
        title = "{A Periodically Active Pulsar Giving Insight into Magnetospheric Physics}",
      journal = {Science},
         year = 2006,
        month = apr,
       volume = {312},
       number = {5773},
        pages = {549-551},
          doi = {10.1126/science.1124060},
archivePrefix = {arXiv},
       eprint = {astro-ph/0604605},
 primaryClass = {astro-ph},
       adsurl = {https://ui.adsabs.harvard.edu/abs/2006Sci...312..549K}
}

@ARTICLE{1986ApJ...301..901R,
       author = {{Rankin}, J.~M.},
        title = "{Toward an Empirical Theory of Pulsar Emission. III. Mode Changing, Drifting Subpulses, and Pulse Nulling}",
      journal = {\apj},
         year = 1986,
        month = feb,
       volume = {301},
        pages = {901},
          doi = {10.1086/163955},
       adsurl = {https://ui.adsabs.harvard.edu/abs/1986ApJ...301..901R}
}

@ARTICLE{2011PASA...28..202H,
       author = {{Hobbs}, G. and {Miller}, D. and {Manchester}, R.~N. and {Dempsey}, J. and {Chapman}, J.~M. and {Khoo}, J. and {Applegate}, J. and {Bailes}, M. and {Bhat}, N.~D.~R. and {Bridle}, R. and {Borg}, A. and {Brown}, A. and {Burnett}, C. and {Camilo}, F. and {Cattalini}, C. and {Chaudhary}, A. and {Chen}, R. and {D'Amico}, N. and {Kedziora-Chudczer}, L. and {Cornwell}, T. and {George}, R. and {Hampson}, G. and {Hepburn}, M. and {Jameson}, A. and {Keith}, M. and {Kelly}, T. and {Kosmynin}, A. and {Lenc}, E. and {Lorimer}, D. and {Love}, C. and {Lyne}, A. and {McIntyre}, V. and {Morrissey}, J. and {Pienaar}, M. and {Reynolds}, J. and {Ryder}, G. and {Sarkissian}, J. and {Stevenson}, A. and {Treloar}, A. and {van Straten}, W. and {Whiting}, M. and {Wilson}, G.},
        title = "{The Parkes Observatory Pulsar Data Archive}",
      journal = {\pasa},
         year = 2011,
        month = aug,
       volume = {28},
       number = {3},
        pages = {202-214},
          doi = {10.1071/AS11016},
archivePrefix = {arXiv},
       eprint = {1105.5746},
 primaryClass = {astro-ph.IM},
       adsurl = {https://ui.adsabs.harvard.edu/abs/2011PASA...28..202H}
}
\bibliographystyle{aasjournal}

\appendix

\section{Removed Observations of PSR~J1745$-$3040}
\label{app:removed}

Several observations of PSR~J1745$-$3040 exhibit anomalous distortions in the pulse-profile baseline relative to the rest of the dataset. In particular, the trough between the minor and main pulse components (centred approximately at phase bin 113) shows a pronounced dip to strongly negative flux-density values that is not present in the rest of the dataset. These baseline distortions artificially increase the inferred NF and exert a disproportionate influence on the fitted NF gradient. All affected observations occur at late epochs and were therefore capable of significantly biasing the long-term trend. These observations were excluded from the primary analysis as shown in Figure~\ref{J1745-3040_removed}.

\begin{figure*}
\centering
\includegraphics[width = 1.0\textwidth]{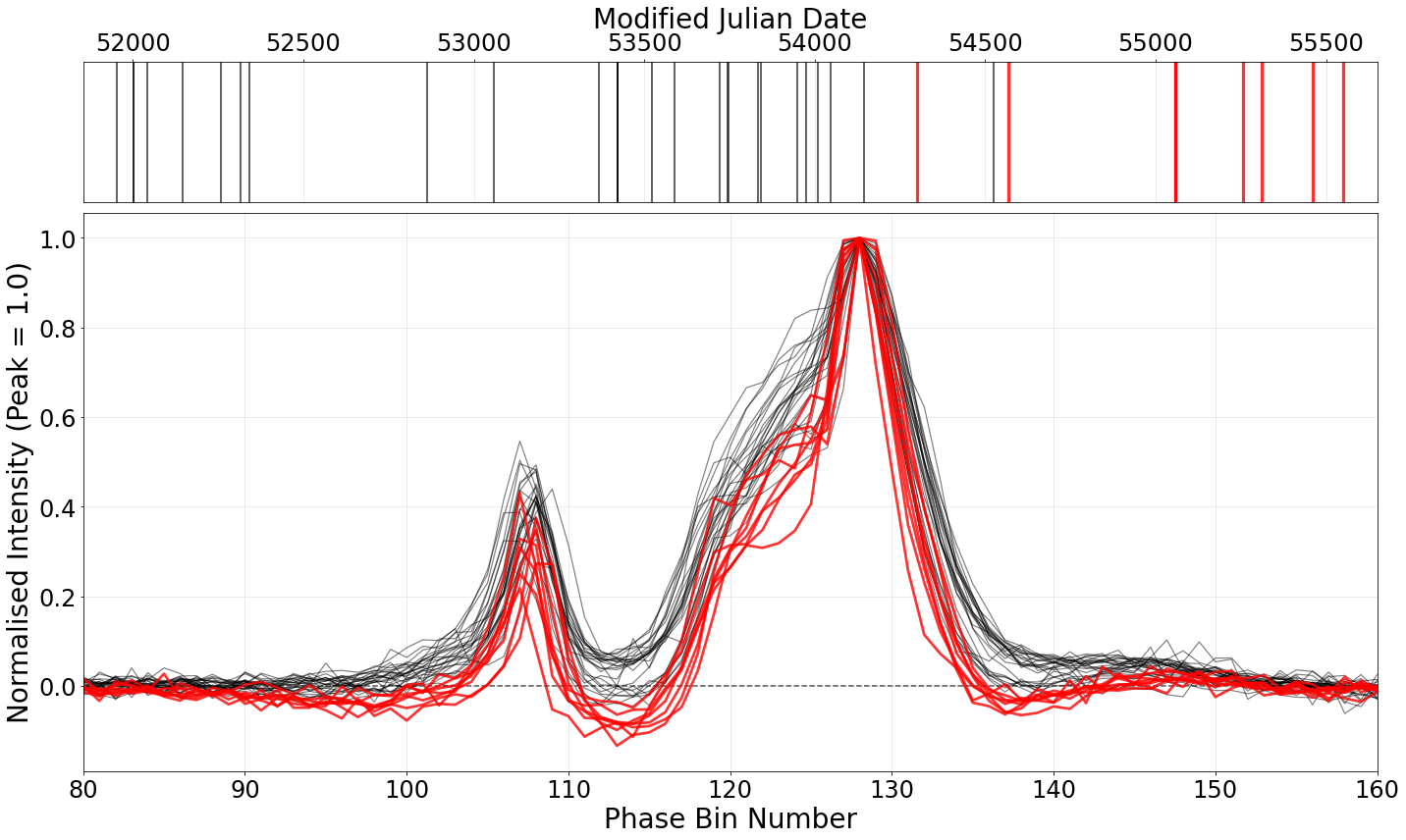}
\caption{Observed pulse profiles of PSR~J1745$-$3040. The upper panel indicates the observation epochs, while the lower panel shows the corresponding pulse profiles. A subset of late-epoch profiles exhibits an anomalously large negative flux-density trough between the minor and main pulse components, centred at approximately phase bin 113. These profiles, and their corresponding observation epochs, are highlighted in red and were excluded from the primary analysis.}
\label{J1745-3040_removed}
\end{figure*}

\end{document}